\documentclass{elsart} \usepackage{graphicx}

\newcommand{\code}	[1]	{\mbox{{\tt #1}}}
\newcommand{\comm}	[1]	{\mbox{{\it #1}}}
\newcommand		{\TE}	{\mbox{$\varepsilon_{{\mathrm{TE}}}$}}
\newcommand		{\TU}[1]{\mbox{$\varepsilon_{#1}$}}

\def\etal			{{et~al.}}
\def\AMRA			{{\sc amra}}
\def\AMRCLAW			{{\sc amrclaw}}
\def\CLOUDY			{{\sc cloudy}}
\def\FISHPAK			{{\sc fishpak}}
\def\FLASH			{{\sc flash}}
\def\GENESIS			{{\sc genesis}}
\def\HERAKLES			{{\sc herakles}}
\def\NIRVANA			{{\sc nirvana}}
\def\PARAMESH			{{\sc paramesh}}
\def\PROM			{{\sc prom}}
\def\PROMETHEUS			{{\sc prometheus}}
\def\RJET			{{\sc rjet}}
\newcommand{\FORTRAN}		{F{\sc ortran~77}}
\def\AA				{A\&A}
\def\CPC			{Comp.\ Phys.\ Comm.}
\def\JCP			{J.\ Comput.\ Phys.}
\def\aupun			{\hbox{$^{\underline{\rm a}}$.}}
\def\Marti			{Mart\'{\i}}
\def\Ibanez			{Ib\'a{\~n}ez}
\begin{document}

\begin{frontmatter}

\title{AMRA: An Adaptive Mesh Refinement Hydrodynamic Code for Astrophysics}

\author{T. Plewa}
\address{
Nicolaus Copernicus Astronomical Center,
Bartycka 18,
00716 Warsaw, Poland
}\ead{tomek@camk.edu.pl}

\author{E. M\"uller\corauthref{ewm}}
\address{
Max-Planck-Institut f\"ur Astrophysik,
Karl-Schwarzschild-Stra{\ss}e 1,
Postfach 1317, 85741 Garching b.\ M\"unchen, Germany
}\ead{emueller@MPA-Garching.MPG.DE}

\corauth[ewm]{Corresponding author.}
\begin{abstract}

Implementation details and test cases of a newly developed
hydrodynamic code, \AMRA, are presented. The numerical scheme exploits
the adaptive mesh refinement technique coupled to modern
high-resolution schemes which are suitable for relativistic and
non-relativistic flows. Various physical processes are incorporated
using the operator splitting approach, and include self-gravity,
nuclear burning, physical viscosity, implicit and explicit schemes for
conductive transport, simplified photoionization, and radiative losses
from an optically thin plasma. Several aspects related to the accuracy
and stability of the scheme are discussed in the context of
hydrodynamic and astrophysical flows.

\end{abstract}


\end{frontmatter}

\section{Introduction}

Many problems in the numerical simulation of hydrodynamic flows
require the use of a high grid resolution in order to describe the
evolution of the system properly. In turn, the use of large numerical
grids implies high computational costs in terms of both, memory and
CPU time. However, in many cases the most important flow features
occupy only a small fraction of the computational domain. Those
structures are usually flow discontinuities like shock waves or
contact surfaces.  The addition of physical processes may lead to the
formation of qualitatively new features which, similarly to flow
discontinuities, can occupy only a small fraction of the total
volume. A proper description of the additional physics may also
require a resolution of time scales which might be much smaller than
the hydrodynamic time scale.

For some problems it is resolution in mass rather than in space or
time which determines the quality of a numerical solution. Here
methods exploiting the Lagrangian rather than Eulerian description of
the flow might be more suitable.  However, the Lagrangian approach
encounters severe difficulties in multidimensional problems due to
large grid distortions typical for shearing flows. Sophisticated grid
rezoning algorithms \cite{cale} did not gain much popularity except
perhaps for one-dimensional problems \cite{sage,titan}. Alternatively,
the notion of a grid might be avoided completely leading to a meshless
code. The Smoothed Particle Hydrodynamics method \cite{l77,gm77,m92}
is a practical example of the realization of this idea. In passing we
note, that in case of discontinuous flows with shocks the use of
artificial viscosity prevents the SPH method from reproducing the
quality of the solutions obtained with modern shock-capturing
advection schemes implemented in the majority of Eulerian codes
\cite{sm93,mm97}.

One may also consider a Cartesian method in which the grid zones are
consecutively refined to increase the resolution where it is desired
from the point of view of numerical accuracy. In this approach the
grid refinement is done on the basis of single cells resulting in a
grid which usually has to be described using a complicated data
structure \cite{mcobra,fr93,dp93,kh98}. When only one isolated region
of the volume has to be refined, an approach with fully nested grids
of increased resolution, similar to multigrid methods with local
refinements for elliptic \cite{b77,b84,bb87} or parabolic problems
\cite{cs73,vlugr3}, might be used \cite{r92,zy97}.

\begin{figure}
\begin{center}
\includegraphics[bb =  30 178 563 666, height=7 cm]{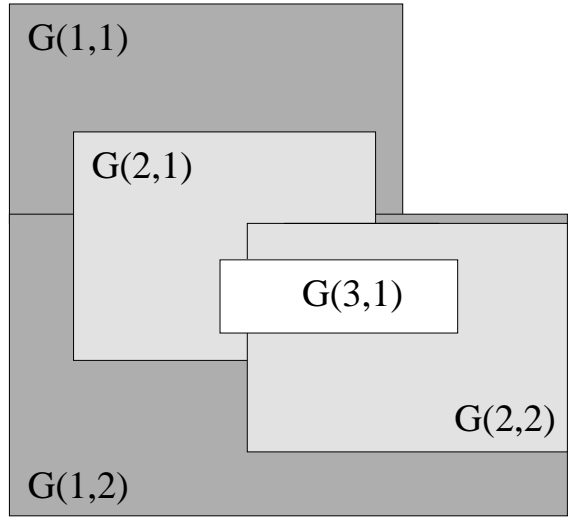}
\caption[]{ 
Hierarchy of grids in Adaptive Mesh Refinement method. The base level
covers the whole computational domain with mesh patches G(1,1) and
G(1,2). Patches G(2,1) and G(2,2) form the second level and a single
mesh patch, G(3,1), is located on the finest level. Notice that the
hierarchy is fully nested with finer mesh patches completely covered
by patches located on the next coarser level, and that each patch may
have more than one parent or/and offspring, and siblings may overlap
each other.
}
\label{f:grids}
\end{center}
\end{figure}

In \AMRA\ we adopt a block-structured approach to grid refinement in
which the refined grid cells are clustered together to form larger
rectangular regions, or mesh patches, overlying parent level
grids. The refinement process is recursively applied to newly created
fine mesh patches in order to increase the resolution even further. In
this way, the final data structure can be seen as a hierarchy of mesh
patches (Fig.~\ref{f:grids})
located at different levels and integrated with individual time steps.
Each single mesh patch has a logical structure identical to the
original numerical grid. This scheme is commonly called adaptive mesh
refinement, AMR, and has been originally proposed by Berger and her
collaborators \cite{bo84,bc89,bell+94,amrclaw}.

\section{Description of the code}			\label{s:method}

\begin{figure}
\begin{center}
\includegraphics[bb =  35 165 560 676, width=0.9\columnwidth]{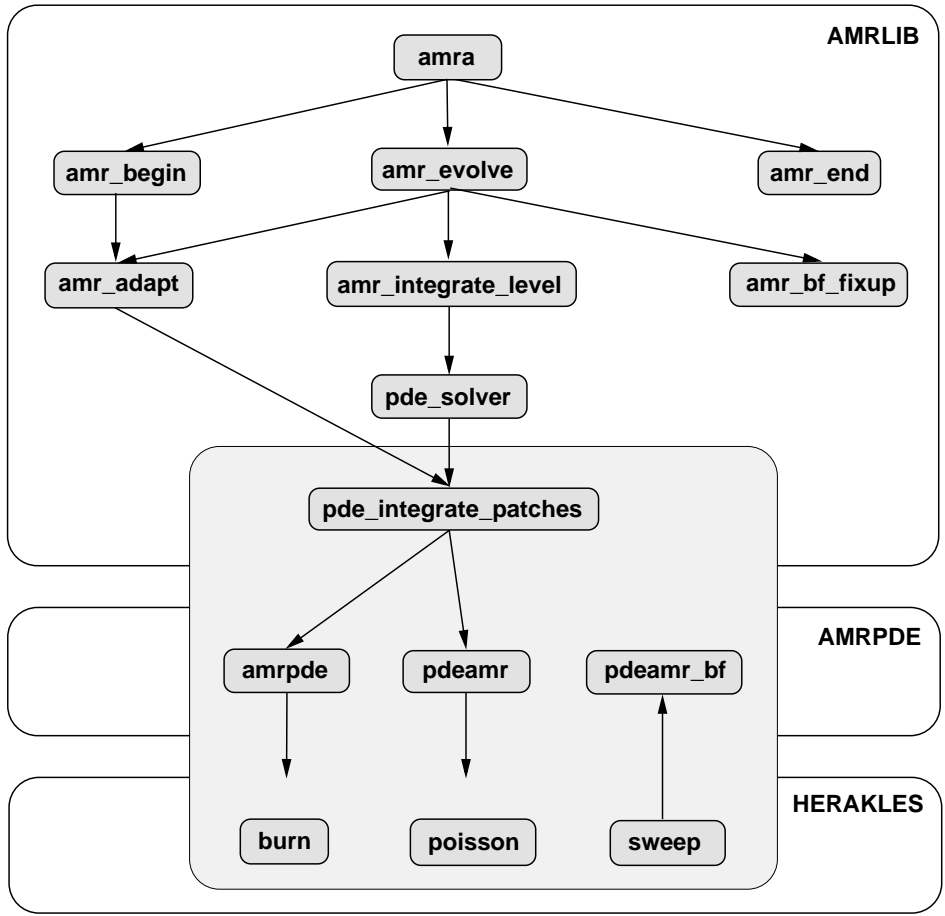}
\caption[]{ 
The basic components of the \AMRA\ code configured for solving
hydrodynamic equations: a library of AMR modules (\code{AMRLIB}); the
user interface (\code{AMRPDE}); the partial differential equations
solver (\code{HERAKLES}). See text for details.
}
\label{f:flow}
\end{center}
\end{figure}

In our description of \AMRA\ we will give an overview of the three
basic components of the code: the library of AMR modules which
orchestrates the execution of the code, the user interface which
serves as a problem independent communication tool between the AMR
driver and the third component of the code -- the partial differential
equations solver (Fig.~\ref{f:flow}).
We will also briefly describe the way in which physical processes are
included in our code -- those which have strictly local character
(radiative losses and nuclear burning) and those which couple
different regions of the computational domain (self-gravity and
thermal diffusion, photoionization). We will conclude our description
with comments on code implementation and code performance on shared
memory parallel machines.

\subsection{Adaptive Mesh Refinement modules}		\label{s:amr}

Our implementation of the AMR algorithm in \AMRA\ closely follows the
description given by Berger and Colella \cite{bc89}. The code is
written in standard \FORTRAN\ and runs on several different
architectures without any modifications. Portability across different
platforms is achieved by defining a set of architecture dependent UNIX
\code{m4} preprocessor directives. Compiler options are automatically
set by a configuration script written in UNIX Bourne shell, and passed
down to the compiler system.

\AMRA\ offers the possibility to couple existing partial differential
equation solvers via a user interface to the AMR modules and its
design is not limited to problems specific for hydrodynamics. All
solvers are supplemented with procedures which handle the
communication with the AMR driver (user interfaces) and a set of
subroutines defining initial conditions for a variety of problems.

\subsubsection{AMR driver}				\label{s:driver}

\begin{figure}
\begin{center}
\includegraphics[bb =  30 268 563 575, width=0.9\columnwidth]{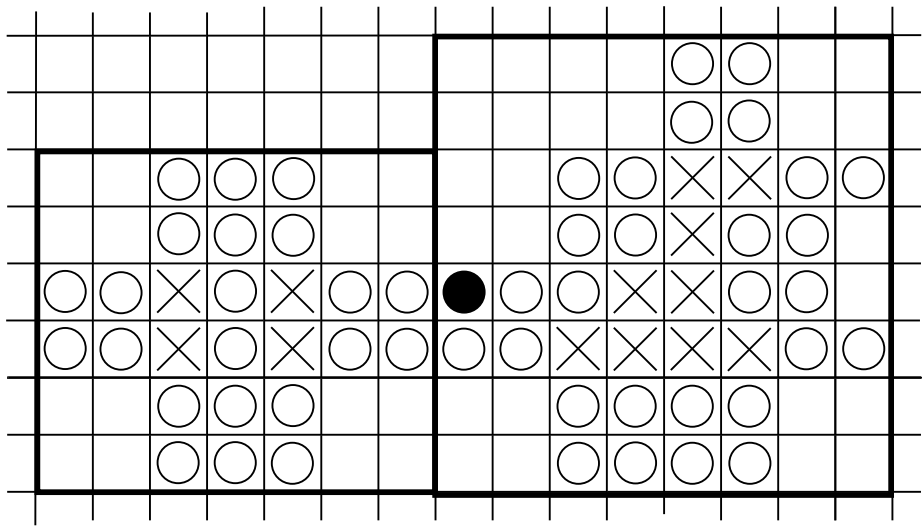}
\caption[]{ 
Box creation in \AMRA. Zones flagged for refinement are marked with a
cross. Each flagged zone is surrounded by (two) buffer zones. Gaps
between flagged or buffer zones smaller than half of the number of
buffer zones are also flagged (zone with a filled circle). Two
boxes are created by the clustering procedure, because the filling
factor of the single rectangle embodying all flagged zones and buffer
zones (63/120) is too small if a filling factor greater than 60\% is
required.
}
\label{f:flags}
\end{center}
\end{figure}

The role of the AMR driver (module \code{amr\_evolve} in
Fig.~\ref{f:flow}) is to initialize and adapt the grid structure
during the simulation and to synchronize the integration process. At
the beginning of the run the user has to specify the (maximum) number
of grid levels and patches for the current run. Since \FORTRAN\ does
not allow for dynamic memory allocation, ultimate upper limits for the
number of levels and patches are defined during the configuration
step. In practice both limits are imposed by the available system
memory. The user also has to define the number of buffer zones (zones
with open circles in Fig.~\ref{f:flags})
to be added around any zone flagged for refinement (zones with crosses
in Fig.~\ref{f:flags}), the minimum filling factor (fraction of
flagged zones) of newly created boxes which become patches the zones
inside the boxes are assigned data. In addition the user has to
specify the order of the conservative multidimensional interpolation
of the state variables at the patch boundaries and from parent patches
to the interiors of boxes created during the adaption step (see
Sect.~\ref{s:adapt}). First, second and third order accurate
interpolation can be used. The dimensionality of the problem and the
type of geometry are defined next. \AMRA\ handles three basic types of
coordinate systems (cartesian, cylindrical, and spherical) in one, two
and three dimensions.

The frequency at which the grid adaption procedure is to be used is given
as a multiple number of parent level steps. If this parameter is set
to zero the hierarchy of levels will remain static after its initial
creation.  The mesh refinement ratios are specified for each grid
level and for each spatial dimension separately. Refinement ratios
might be different in every coordinate direction, can take arbitrary
integer values greater than zero, and must remain constant during the
whole run.  Additional temporal refinement can be specified separately
for each level. In this case the integrator subcycles over the nominal
time step. Finally, the order of the spatial accuracy of the solver
(or type of the solver) can also be specified separately for each
level.

\subsubsection{Code operation}				\label{s:exec}

Execution of the code begins with the initialization of various
counters and system dependent constants followed by opening files
describing actual control parameters of the simulation. Immediately
after reading input parameters, the code checks for their consistency
with the code configuration (e.g., internal dimensioning of
arrays). Next the initial hierarchy of levels is created starting from
the base level until the maximum number of levels is reached or when
all of the refinement criteria are fulfilled. The initial model is
printed out and the execution enters the main integration loop.

\begin{figure}
\begin{center}
\includegraphics[bb =  35 343 560 499, width=0.9\columnwidth]{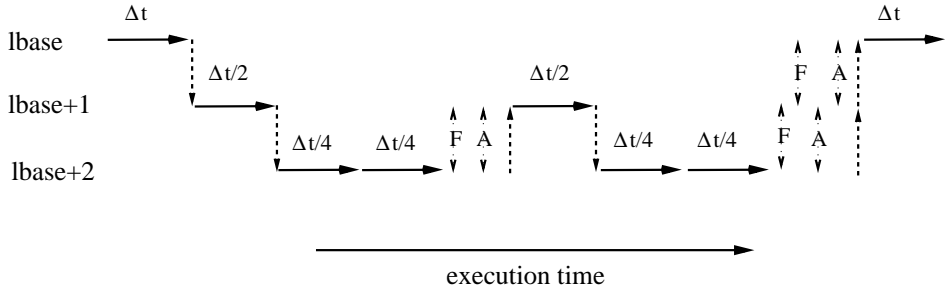}
\caption[]{ 
Execution cycle during a single base level step of \AMRA. Three levels
of grids are used the refinement factor being two between levels. Grid
adaption and flux fixup steps are denoted by ``A'' and ``F'',
respectively.
}
\label{f:cycle}
\end{center}
\end{figure}

During a time step, grid levels are integrated recursively (module
\code{amr\_evolve} in Fig.~\ref{f:flow}) in a way which resembles a
V-cycle of a multigrid solver. The execution sequence is presented in
Fig.~\ref{f:cycle}.
Starting at the base level the equations are integrated for a full
time step (modules \code{amr\_integrate\_level}, \code{pde\_solver}, and
\code{pde\_integrate\_patches} in Fig.~\ref{f:flow}). On fine levels the
solution is advanced for a single step (if there are still finer
levels) or for a number of steps necessary to reach the evolutionary
time of the parent level.

The sequence of operations during integration of a single patch is as
follows: Just prior to a patch integration its current state is saved
in order to preserve the data required for the temporal interpolation
of the state on the child patches at their fine-coarse boundaries (see
Sect.~\ref{s:adapt}).\footnote{In this respect \AMRA\ more closely
follows the prescription of \cite{bc89} than, for example, \NIRVANA\
or \PARAMESH\ and {\FLASH}, for which no temporal interpolation of the
state is done at the fine-coarse boundaries. In these codes all the
levels in the hierarchy are integrated with a common time step size.}
Also, the source terms and the boundary fluxes to be summed over a
time step (see below) are initialized. The latter operation is
necessary only at the boundaries with child patches and, therefore, it
is not required for patches located on the finest level. Next boundary
data is provided for each patch, and patches are integrated until the
time of the parent level is reached. After integration of the patch is
completed, fluxes are fixed up (``F'' operator in
Fig.~\ref{f:cycle}). The fixup step is necessary to account for the
difference between the flow across the boundary between the child
patch and the corresponding zone interface of the parent patch
(fine-coarse boundary in Fig.~\ref{f:bounds}). This is to ensure
global conservation of the advection process (see \cite{bc89} for
details). Finally, the solution obtained on a fine patch is projected
up to its parent patches. If desired, an adaption step (``A'' operator
in Fig.~\ref{f:cycle}) can follow.

The above procedure is repeated until the finest level is advanced to
the time of the base level, i.e. until all grid levels are
synchronized.  The size for the next time step is calculated as
follows. Firstly, the limiting time step is computed for each level by
taking the global minimum over all patches located on that
level. Next, these level-dependent time steps are appropriately scaled
to account for the change in spatial resolution between levels. The
time step is then given by the minimum of these scaled level-dependent
time steps. Eventually code execution is terminated with a call to
subroutine \code{amr\_end} (see Fig.~\ref{f:flow}) which prints
statistics about the CPU time used by different parts of the code, the
number of zones evolved on each level and the estimated speedup. It
also dumps a restart file and ensures proper closing of output files
opened during run.

The execution of \AMRA\ can be modified during runtime using
``messages'' which are read from file \code{00\_message}. A sin\-gle
line of this file contains blank-separated three character strings:
\code{'amra'}, \code{'cmd'} and \code{'s'}. Here \code{cmd} is a
unique string of characters recognized by the code (command) and the
single character \code{'s'} can be \code{'+'} (command \code{cmd} is
activated) or \code{'-'} (command is deactivated). Commands allow to
observe the current code progress tracing most important (integration,
adaption) subroutine calls, dumping restart or checkpoint (with an
only partially synchronized state between levels) files, and stopping
(after full time step) or smoothly terminating (after level
integration) code execution. Independently of the above mechanism,
minimal images of the code memory (required for restart) are written
in predefined intervals of wall-clock time throughout code execution.

\subsubsection{Adaption of grid hierarchy}			\label{s:adapt}

The creation of patches comprises four independent stages (module
\code{amr\_adapt} in Fig.~\ref{f:flow}):
\begin{description}
\item[flagging:] identification of regions on the current (coarse)
level which need to be resolved at higher resolution (fine level);
\item[clustering:] definition of boxes in a way that their set
completely covers all regions identified in the previous stage;
\item[optimization:] merging or splitting of boxes aimed at obtaining
better code performance;
\item[data assignment:] filling the boxes obtained in the optimization
step with necessary data.
\end{description}

\paragraph{Flagging for refinement}			\label{s:flag}

\AMRA\ offers three independent ways to identify (flag) regions which
have to be refined. An estimate of the local truncation error of the
solver can be obtained by comparing solutions obtained at the nominal
and at a twofold lower resolution (for which the data is obtained by
taking an appropriate average over the state on the original patch)
after evolving both states for two nominal time steps
\cite{bo84,oz96}. Zones are flagged for refinement once the estimate
obtained that way exceeds a certain threshold, {\TE}. No truncation
error can be estimated when one of the refinement ratios is odd.

A much simpler method of ``error forecasting'' (which is also much
cheaper since it does not require additional calls to the solver) is
based on the local relative change of the values of selected
quantities (e.g., the gas density or the pressure). The justification
for this approach relies on the observation that most of the local
variation of the function is contained in its first derivative (first
order term in Taylor expansion). One can modify this approach by
introducing additional resolution-dependent scaling
\cite{nirvana}. Since a small value of the second derivative does not
guarantee that the gradient is small, for solvers based on higher
order schemes one may supplement (rather than replace) the above
procedure by introducing non-dimensional error indicators based on the
first and second spatial derivatives of the state variables
\cite{l95,flash}. Finally, \AMRA\ allows for direct modification of
error flags by the user through the user interface (see
Sect.~\ref{s:ui}).

The mesh generation algorithm uses integer flags which are set by the
flagging module. Buffer zones are added around regions flagged for
refinement in order to prevent discontinuities from escaping from fine
patches during the following integration process (zones with open
circles in Fig.~\ref{f:flags}). In addition, regions which separate
flagged or buffer zones and which are smaller than half the buffer
length are also marked for refinement (zone with a filled circle in
Fig.~\ref{f:flags}). Finally, proper nesting is ensured by flagging
zones in the regions occupied by patches on the next but one finer
level. We note that the use of integer instead of logical flags is
necessary to distinguish ``error'' and ``buffer'' regions in those
cases where the continuity of the grid hierarchy is essential (e.g.,
when using periodic boundary conditions).

\paragraph{Clustering}					\label{s:cluster}

The clustering procedure returns a set of boxes (i.e., unfilled
patches) identified by the positions of their corners in physical
space (rectangles drawn with thick lines in Fig.~\ref{f:flags}). The
complete set of boxes totally covers the flagged zones subject to two
constraints: 1) the ratio of flagged to total box volume (box
efficiency) must not be smaller than some specified threshold (e.g.,
60\%); 2) the set needs to be completely embedded in coarse
patches. An initial distribution of boxes is created either by using a
simple method of bisection \cite{bo84} or with a more advanced point
clustering algorithm \cite{br91}. For the example shown in
Fig.~\ref{f:flags} two boxes are created by the clustering procedure,
because the filling factor of the single rectangle embodying all
flagged zones and buffer zones (63/120) is too small, if a filling
factor larger than 60\% is required.

\paragraph{Optimization of grid hierarchy}		\label{s:optim}

The distribution of boxes resulting from the clustering stage can be
optimized by an optional merging/splitting step (this step may create
partially overlapping sibling patches).  This additional optimization
is aimed at reducing the overall execution time and exploits
information about the current computer architecture (see
Sect.~\ref{s:parallel}) and the instantaneous (i.e., for the present
hierarchy of grids) AMR overhead due to the inter-grid-level
operations.

\paragraph{Data assignment}				\label{s:dassign}

\begin{figure}
\begin{center}
\includegraphics[bb =  35 217 561 625, height=7 cm]{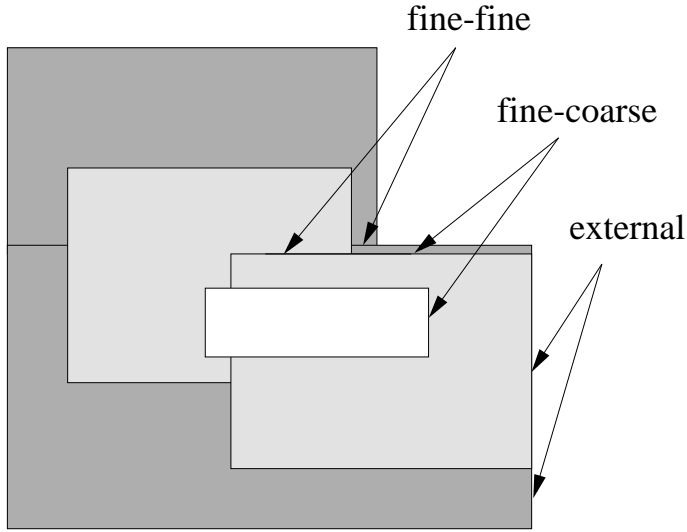}
\caption[]{ 
Types of boundary zones in AMR hierarchy.
}
\label{f:bounds}
\end{center}
\end{figure}

During the final stage of the adaption step, free memory entries are
identified and reserved for new patches. Existing patches may
eventually be shifted up or down in the running patch index space in
order to keep the memory contiguously filled. The geometry of new
patches is defined by the geometry of the boxes which were created
during the clustering step and the relationships between patches and their
neighbours and parents are determined. The initial state for the patch
interior is provided by a simple copy operation of existing fine data,
or by conservative interpolation from parent patches. Boundary (ghost)
zones can be tagged as external (located outside the computational
domain), fine-fine (overlapped by the interior zones of a sibling
patch), or fine-coarse (overlapped by the interior of a coarse, not
necessarily parent, patch). These three types of boundary zones are
depicted in Fig.~\ref{f:bounds}.
Data for ghost zones tagged as fine-coarse is obtained by conservative
interpolation \cite{dm00}.  Notice that the boundary type is
characteristic of the individual ghost zone as it may change along the
patch boundary (see upper boundary of the lower right mesh located on the
second level in Fig.~\ref{f:bounds}).

\subsubsection{\AMRA\ data flow and structure}		\label{s:data}

Since \AMRA\ is exclusively written in \FORTRAN\ the data structure is
static in nature, with the maximum number of patches and their sizes
being determined during the compilation step. It has to be noted, that
the static memory assignment used in \AMRA\ does not necessarily imply
a severe waste of memory since the clustering procedure guarantees
that patches are always densely filled with flagged zones (see
Sect.~\ref{s:adapt}).

The data structure of \AMRA\ consists of three different parts: the
main data block used for storing the hydrodynamic state and the source
terms, a set of pointers describing the grid structure, and an
additional workspace used by \AMRA\ for mesh generation.

The hydrodynamic state and the source terms are kept in two
5-dimensional arrays: \code{STATE(i,j,k,ius,jg)} and
\code{SOURCE(i,j,k,iso,jg)}. The triple of indices (\code{i,j,k})
corresponds to the spatial location of the zone within the patch,
\code{ius} (\code{iso}) denotes state variables (source terms), and
\code{jg} is the patch index. There are two additional 5-D arrays
which contain the solution and sources obtained at the previous time
step. If necessary, these can be used for temporal interpolation of
boundary data or extrapolation of source terms.

For each corner of each patch a pointer is used for the inter-patch
communication. The value of the pointer identifies the zone on the
sibling or parent patch overlapping the corner zone. Each patch has
additional attributes like the level on which the patch is located,
the number of parent and child patches, and a set of pointers to the
parents and offsprings. Grid levels have attributes, too. They are
required for performing global operations on patches, levels, or the
whole domain. The attributes are the number of mesh patches occupying
the level and a pointer to the last patch on the given level.
Finally, since the code provides an option for mesh flagging with the
truncation error estimation procedure, \AMRA\ uses temporary work
storage to preserve the original state of the patch for which the
truncation error is currently estimated.

In other \FORTRAN\ implementations of AMR a memory buffer is allocated
with a size equal to the maximum memory partition available on a
specific system. This approach has been used by Quirk \cite{q91} and
Berger and LeVeque \cite{amrclaw} in \AMRCLAW.  However, using a
memory buffer, i.e. a linear address space, adds a certain complexity
to the code since each access to a specific portion of the data has to
be calculated explicitly. In the approach adopted by MacNeice \etal\
\cite{paramesh} in \PARAMESH\ (which provides AMR functionality also
for the \FLASH\ code \cite{flash}) and by Ziegler in \NIRVANA\
\cite{nirvana} patches are of a fixed size in each coordinate
direction (8 zones in case of \PARAMESH\ and 4 zones in
\NIRVANA). Small patch sizes have the advantage of making the adaption
process very flexible and effective (patches are always densely filled
with flagged zones) but are disadvantageous when higher order schemes,
which require long stencils, are used.  In this case, the number of
ghost zones might become comparable to the number of active
zones.\footnote{For a solver requiring 4 ghost zones (e.g., a PPM
scheme) only $(4/12)^3\approx 4$\% (\NIRVANA) or $(8/16)^3=12.5$\%
(\PARAMESH) of the allocated memory would be effectively used by the
solver.} On the other hand, a patch of small size requires larger
amount of operations at its boundaries relative to the integration of
the equations. Finally, fine grained computations diminish code
performance especially on machines with vector architecture (see
Sect.~\ref{s:parallel}).

The usage of computer memory by \AMRA\ is defined at the level of
patch creation. Table~\ref{t:mscaling}
\begin{table}
\caption[]
{
Dependence of \AMRA\ memory size on the number of levels (\code{mlg}) and
patches (\code{mgg}) given as a function of patch size for a state
vector of length 8 (i.e, a 3-D flow with density, momenta,
total energy, and three additional arrays used as work storage).
}
\label{t:mscaling}
\begin{flushleft}
\begin{tabular}{lrrr|lrrr}
\hline
\noalign{\smallskip}
\multicolumn{4}{c}{2-D} \vline & \multicolumn{4}{c}{3-D} \\
\noalign{\smallskip}
\hline
\noalign{\smallskip}
patch size & \code{mlg} & \code{mgg} &  memory &patch size & \code{mlg} & \code{mgg} &  memory \\
\noalign{\smallskip}
           &            &            & [Mbytes] &           &            &            &  [Mbytes]\\
\noalign{\smallskip}
\hline
\noalign{\smallskip}
$64^2$  &      1        &      1      &       3  &  $8^3$  &      1        &      1      &       3   \\
$64^2$  &      1        &    100      &     109  &  $8^3$  &      1        &    100      &     105   \\
$64^2$  &      2        &    100      &     124  &  $8^3$  &      2        &    100      &     118   \\
$64^2$  &      1        &   1000      &    1091  &  $8^3$  &      1        &   1000      &    1037   \\
$64^2$  &      2        &   1000      &    1355  &  $8^3$  &      2        &   1000      &    1296   \\
$64^2$  &     10        &   1000      &    1355  &  $8^3$  &     10        &   1000      &    1296   \\
$128^2$ &     10        &    100      &     425  &  $16^3$ &      1        &    100      &     335   \\
$128^2$ &     10        &    500      &    2117  &  $16^3$ &      2        &    100      &     381   \\
$128^2$ &     10        &   1000      &    4308  &  $16^3$ &     10        &    100      &     381   \\
        &               &             &          &  $16^3$ &     10        &    500      &    1902   \\
        &               &             &          &  $16^3$ &     10        &   1000      &    3879   \\
\noalign{\smallskip}
\hline
\end{tabular}
\end{flushleft}
\end{table}
presents the actual memory required by \AMRA\ for different maximum
number of levels (\code{mlg}) and patches (\code{mgg}) in two and
three dimensions. It can be seen that the memory usage scales almost
linearly with the number of patches, and that it does not depend on
the number of levels;\footnote{Since there is no formal limitation
regarding the number of levels their number might be arbitrarily large
although in practice no more than 10 levels are usually used.}
neither does the linear dependence of memory size on the number of
patches change with patch size. Comparing memory use for single level
(non-AMR) and two-level (with the support for AMR compiled in)
configurations we estimate that the memory overhead caused by AMR
varies between $\sim 15$\% (for $\code{mgg}=100$) to about $\sim 30$\%
($\code{mgg}=1000$).

\subsection{User interface}				\label{s:ui}

The design of the user interface is crucial for the ease with which a
new problem can be set up. The number of places which have to be
modified for a new problem should be minimized and all portions of the
code independent of the problem have to be well separated. The AMR
data structure, the grid generator, the interpolation of boundary
data, and the recursive process of integration, are all handled by
the AMR modules. These and the major part of the user interface do not
need to be modified for a new problem.

The user interface consists of several subroutines which allow for
proper communication and data flow between the AMR modules and the
partial differential equations (PDE) solver. The primary role of the
user interface is to perform the necessary data copy operations from
and to AMR data storage (modules \code{amrpde} and \code{pdeamr} in
Fig.~\ref{f:flow}) and to provide external boundary conditions. The
user interface contains fully implemented reflecting, transmitting,
and periodic boundary conditions for the hydrodynamic state, and
boundary conditions for the gravitational potential. Also, the user
interface is responsible for keeping a record of numerical fluxes
calculated by the solver, which are stored in flux counters and used
later during flux fixup step (modules \code{pdeamr\_bf} and
\code{amr\_bf\_fixup} in Fig.~\ref{f:flow}).

In addition, the user interface contains a subroutine which allows for
a direct modification of the flags used during the adaption step (see
Sect.~\ref{s:adapt}). The extent of the base level can also be
modified by adding or removing patches provided that the state on all
levels is synchronized in time. Special care must be paid to save all
problem dependent quantities which have to be restored upon
restart. It is also possible to replace any standard output routine
provided with \AMRA\ to create customized output. Several standard
equations of state can be selected during the configuration step:
ideal or isothermal gas, a fully ionized plasma of arbitrary chemical
composition, a mixture of Boltzmann gases with radiation, or a nuclear
equation of state.

\subsection{Solvers overview}				\label{s:solvers}

Although \AMRA\ can be used for any numerical problem which requires
discretization of the underlying equations, so far it has been used
only for the solution of hyperbolic partial differential equations
(PDE) specific for hydrodynamics and in application to astrophysical
problems. The current version of \AMRA\ includes two implementations
of the PPM method \cite{cw84}, the \PROMETHEUS\ \cite{fma89} and
\HERAKLES\ \cite{herakles} codes, the special relativistic solver
\RJET\ \cite{rjet} and its modern version \GENESIS\ \cite{genesis}.

The PDE solver (module \code{pde\_solver} in Fig.~\ref{f:flow}) is a
workhorse for \AMRA, and much effort has been put into linking
existing user codes to the AMR part as easy as possible. Usually, the
adaption of a new solver begins with the removal of unnecessary output
operations which are completely handled by AMR modules or the user
interface. Problem dependent modules have to be separated and, if
necessary, any custom or defined boundary condition have to be moved to the
user interface. The existing main program has to be converted into a
stand-alone subroutine which has to solve the evolutionary equations
for a single time step and to return the boundary fluxes. Finally,
care has to be taken as to remove any assumed dependencies between
zone numbering and their physical coordinates.

The \PROM\ solver is based on \PROMETHEUS, a multidimensional
implementation of the Direct Eulerian Piecewise Parabolic Method of
Colella and Woodward \cite{cw84}, originally developed by Fryxell,
M\"uller and Arnett \cite{fma89,m98}. With respect to \PROMETHEUS,
\PROM\ differs in details of the calculation of the effective states,
the solution of the Riemann problem, the dissipation mechanisms
(flattening and artificial viscosity modules), conservative angular
momentum transport, and adaption to a rotating frame of reference with
conservative treatment of the Coriolis force \cite{k98}. A more
substantial modification is the inclusion of the CMA method \cite{cma}
for multifluid advection. The physics already included in \PROMETHEUS\
(self-gravity, nuclear burning, realistic equation of state
\cite{cg85}) has been extended by radiative cooling, thermal diffusion
and conduction, and photoionization. For parallel implementation on
shared memory machines, most of the solver memory has been declared as
private with the exception of the input configuration parameters
required by {\PROMETHEUS}. Finally, the poor performance of \PROM\ on
small grids on machines with vector architecture resulted in its
complete rewrite and creation of the \HERAKLES\ solver. We defer a
detailed description of this new solver to a forthcoming publication
\cite{herakles} (see also Sect.~\ref{s:perfopt}).

For problems involving relativistic flows we adapted the \RJET\ solver
and the \GENESIS\ solver. Since both codes use a multi-staged (second
or third order) Runge-Kutta integrator for advancing the solution in
time, intermediate fluxes have to be stored for each patch in AMR
memory. These are updated with partial fluxes after each stage of the
integrator, and are passed to AMR after the last stage of the integration
process. Except for this, the coupling of \RJET\ and \GENESIS\ to AMR
library required similar modifications as in the case of \PROMETHEUS.

\subsection{Problem set-up}				\label{s:problem}

Initial conditions are defined with the help of a set of subroutines
stored in a single file.  The basic configuration of \AMRA\ (maximum
number of levels and patches, output type) and the configuration of
the solver (solver type, problem code name, maximum patch size,
external boundary conditions, number of fluids, type of equation of
state, physics options) are defined by a set of UNIX \code{m4}
preprocessor directives declared in a problem configuration
file. Table~\ref{t:problem}
\begin{table}
\caption[]
{
\AMRA\ problem configuration file for the Hawley-Zabusky test problem
(see Sect.~\ref{s:hz}).
}
\label{t:problem}
\begin{flushleft}
\begin{tabular}{ll}
\hline
\noalign{\smallskip}
\code{m4} definition             & comment \\
\noalign{\smallskip}
\hline
\noalign{\smallskip}
\code{define(PROBLEM,HZ)}        &\comm{problem code name} \\
\code{define(MGG,400)}		 &\comm{maximum number of patches} \\
\code{define(MLG,3)}		 &\comm{maximum number of levels} \\
\code{define(MX1G,60)}		 &\comm{maximum patch size in 1st dimension} \\
\code{define(MX2G,10)}           &\comm{maximum patch size in 2nd dimension} \\
\code{define(MGZG,4)}		 &\comm{number of ghost zones} \\
\code{define(GAS,1)}		 &\comm{number of gaseous components} \\
\code{define(NFLUID\_G,1)}	 &\comm{number of gaseous fluids} \\
\code{define(NCONSERVED,5)}	 &\comm{number of conserved variables} \\
\code{define(NEXTRA,3)}		 &\comm{number of temporary variables} \\
\code{define(NSOURCES,0)}	 &\comm{number of source terms} \\
\code{define(BC\_L1,UGBC)}	 &\comm{boundary type, left edge, 1st dimension: inflow} \\
\code{define(BC\_R1,TR)}         &\comm{boundary type, right edge, 1st dimension: transmitting} \\
\code{define(BC\_L2,RE)}         &\comm{boundary type, left edge, 2nd dimension: reflecting} \\
\code{define(BC\_R2,RE)}         &\comm{boundary type, right edge, 2nd dimension: reflecting} \\
\code{define(EOS,IDEAL)}	 &\comm{use ideal equation of state} \\
\noalign{\smallskip}
\hline
\end{tabular}
\end{flushleft}
\end{table}
shows a problem configuration file for the Hawley-Zabusky test (see
Sect.~\ref{s:hz}). Additional UNIX \code{make} system targets defined
in the main \code{makefile} help saving and restoring particular
problem set-up and input files required during runtime.

The subroutine in which the initial state is defined takes as its
argument the patch level and the patch number, and calculates
geometrical terms for the given geometry type and patch extent in
physical space. The initial state is defined by assigning data
directly to the AMR arrays.  If the multifluid option is used, the
total density is calculated as the sum over partial densities. In the
multifluid case care also has to be taken to advect individual species
consistently with the total density \cite{cma}. The set-up of the
hydrodynamic state is complete after a call to the equation of state.

\subsection{Treatment of physical processes}		\label{s:physics}

Physical processes are treated in \AMRA\ with the help of the operator
splitting technique. In this approach it is assumed that different
processes can be treated independently from each another. In other
words, it is implicitly assumed that a coupling between a given
physical process and the advection occurs on a time scale which is
shorter than the hydrodynamic time scale, and special care must be
taken in cases when both time scales become comparable. A typical
example of such a situation is heat diffusion (e.g., by thermal
conduction or radiation) for which the time scale depends on the
inverse square of the zone width. It can easily be much shorter than
the limit imposed on the time step by the Courant-Friedrichs-Lewy
\cite{cfl} condition. In this case an implicit scheme for the
calculation of thermal energy transport should be used.  However, this
guarantees only that the diffusion process itself will be calculated
correctly (that is, the problem of coupling between the advection and
energy diffusion processes still persists). In passing we note that
the order in which processes are executed in the sequence of operators
is not arbitrary in the stiff case, and in order to minimize errors
arising from the operator splitting the stiffest operators should be
applied last \cite{s00}.

\paragraph{Radiative cooling}				\label{s:cooling}

Radiative cooling for an optically thin plasma is calculated
explicitly for each zone with integration over hydro-timestep done in
small substeps whose length is constrained by the maximum allowed
change in the gas energy \cite{p93}. We note that this approach is
suitable for an arbitrary (i.e., also nonmonotonic) dependence of the
emissivity on temperature. The code can be configured to calculate the
plasma emissivity assuming equilibrium conditions for solar
metallicity \cite{sd93} or with additional metallicity dependence
({\CLOUDY}~90.01, \cite{cloudy}). In addition two nonequilibrium
cooling curves for several \cite[Raymond and Smith code]{s96} or only
solar \cite{sd93} metallicity are available.  The radiative cooling
module is called after the advection step and is followed by an
optional call to the photoionization module (see below).

\paragraph{Nuclear burning}				\label{s:burning}

Similarly to radiative cooling nuclear burning is a purely local
process leading to a modification of the chemical composition (and
energy release) in sufficiently dense and hot regions of the
computational domain abundant in nuclear fuel. Although \AMRA\ allows
for simultaneous conservative transport of an arbitrary number of
fluids representing different nuclear species the use of relatively
large (number of species $n_X > 50$) nuclear reaction networks
(especially in multidimensional calculations) still appears to be
beyond the reach of current computer installations. For stellar
applications, e.g. nova outbursts \cite{kht98} or the early phases of
the shock propagation during a supernova explosion \cite{k+00}), small
networks give adequately accurate results with uncertain reaction
rates, numerical diffusion \cite{cma}, or even inadequate refinement
criteria used in AMR simulations \cite{kpm98} being the dominant
sources of errors in the final chemical composition.

In the present implementation \cite{kpm98} the burning module (module
\code{burn} in Fig.~\ref{f:flow}) solves an $\alpha$-network with 27
reactions. The reactions couple 13 nuclei (\nuc{4}{He}, \nuc{12}{C},
\nuc{16}{O}, \nuc{20}{Ne}, \nuc{24}{Mg}, \nuc{28}{Si}, \nuc{32}{S},
\nuc{36}{Ar}, \nuc{40}{Ca}, \nuc{44}{Ti}, \nuc{48}{Cr}, \nuc{52}{Fe},
\nuc{56}{Ni}). A solution of the coupled nonlinear system of equations
describing the simultaneous evolution of composition and temperature
is obtained implicitly with a Newton-Raphson iteration \cite{m86}. The
energy released during nucleosynthesis is accounted for in the energy
equation. Note that since each chemical element is treated as a
separate state variable, it requires allocation of additional
memory. For the present case of 13 nuclei entries in
Table~\ref{t:mscaling} should be multiplied by a factor $(8+13)/8
\approx 2.6$

\paragraph{Physical viscosity}				\label{s:viscosity}

The viscous stress tensor and divergence components required for the
viscosity terms \cite{k99} are calculated at the beginning of the time
step (at the beginning of the first sweep if directional splitting is
used). In the PPM method viscous forces contribute to the effective
states \cite[Eq.~3.7]{cw84} and are also included in the acceleration
part of the advection step \cite[Eq.~3.8]{cw84}. In passing we note
that since updating momenta in the acceleration step of PPM requires
knowledge of the forces at the end of the time step, the consistent
implementation of viscous forces would make the scheme implicit.

\paragraph{Thermal energy transport}			\label{s:transport}

Energy transport in \AMRA\ includes the processes of thermal
conduction \cite{w44II} and diffusion. Both processes can be treated
explicitly, the energy transport being included in the advection
step. The time step has to be globally reduced if the energy change
due to diffusion exceeds some threshold in valid (not further refined)
regions.

In case the thermal diffusion time scale becomes much shorter than the
hydrodynamic time scale an implicit approach must be used. The
equations are discretized on a ``transport grid'' which has a
resolution equal to the finest resolution in the simulation. At the
beginning of each base level time step, conservative multidimensional
interpolation (also used in the AMR part) of appropriate state
quantities (density, chemical or nuclear composition, internal energy)
is used to provide data on the transport grid. An implicit solution to
the nonlinear diffusion equation is obtained with the help of the
fractional steps method \cite{y71,w44II}. Subsequently the internal
energy is mapped back to each patch.

\paragraph{Photoionization}				\label{s:photo}

Photoionization consists of calculating the local Str\"omgren radius
along the radial direction from the source of photoionizing photons.
Photoionization is included in 1- and 2-D versions of the code. The
central time-dependent source of photoionizing photons can be
specified. Calculation of the hydrogen column density is done on the
transport grid. In spherical geometry the calculation is
straightforward, while in cylindrical geometry ray-tracing is
used. The photoionization procedure returns a map (i.e., a 1-D or 2-D
array depending on the dimensionality of the problem) of ionized
regions which is interpolated for each patch. This information is used
later by the radiative cooling module to modify the gas temperature.

\paragraph{Gravitational forces}			\label{s:gravity}

The simplest source of gravitational acceleration in \AMRA\ are
time-dependent point-like sources arbitrarily distributed in space.
Self-gravity of the gas is calculated with the transport grid
approach. The solution of the Poisson equation (module \code{poisson}
in Fig.~\ref{f:flow}) is obtained by a direct summation of the
contributions of the gas shells (if a spherically symmetric
distribution is assumed) or with help of the \FISHPAK\ FFT library
\cite{fishpak}. In the approach exploiting the transport grid, the
accuracy of the solution is improved by a linear extrapolation of the
potential obtained at the two most recent epochs of the grid
synchronization. However, since the above approach is inefficient in
terms of memory consumption and as the temporal accuracy of the
solution does not allow for studying self-gravitating systems with
adequate accuracy, we plan on implementing a method based on fast
direct solvers \cite{ellds} or a multigrid solver \cite{ellmg}.

\subsection{Data visualization}				\label{s:vis}

For storing the results of the simulations, \AMRA\ provides three
independent data formats: \code{OUT}, \code{PIX}, and \code{MOV}.

In one dimension the relatively small size of data allows for all
output to be written in ASCII format. Files in \code{OUT} format
contain the geometry and the state for all patches and include ghost
zones. \code{MOV} files contain the same data but only for valid
regions (i.e., those regions which are not further refined) . There is
no provision for \code{PIX} output in one dimension, but in case of
need for special output the user has an option for providing a
suitable subroutine.

Two- and three-dimensional data come in variety of formats depending
on visualization tools available for data analysis. In case of
multidimensional simulations data is always stored in binary
format. On some systems conversion between different internal
representations of data (little and big endian) is done via compiler
(Portland Group compilers under Linux) or preprocessor (Cray and SGI)
options with calls to the appropriate system subroutines.

Files in \code{PIX} format are best suited for a quick preview of the
simulation progress. These are images (or voxels in three dimensions)
covering a part or the whole computational domain with a resolution
predefined during the configuration step. \code{OUT} files can be
written in a format suitable for postprocessing with tools like IDL
\cite{k+00} or AVS/Express \cite{pdisk}, and contain all the
information about the hydrodynamic state and the current grid
structure. If more frequent output is required, selected variables can
be written in \code{MOV} format (native binary format) or using the
HDF library \cite{hdf}. Optionally, any of the \AMRA\ output
subroutines can be replaced by appropriate calls to user supplied
subroutines.

In addition, the code distribution contains stand-alone programs and
scripts helping to start with data processing or visualization:
\code{amra\_conv} converts \code{OUT} files to a format suitable for
visualization with AVS/Express (VISA application \cite{pdisk} or
library of modules \cite{fwf98}); \code{avs2pix} converts \code{OUT}
files to \code{PIX} files; \code{idl} is a small library of IDL
subprograms used for visualization of \code{PIX} files.

\subsection{Parallel implementation}			\label{s:parallel}

In its current version \AMRA\ can be executed in parallel on shared
memory systems with vendor-specific autotasking directives or the
OpenMP standard \cite{openmp}. The code has been used with success on
Cray Parallel Vector Processor (PVP) systems (Y-MP, J90 and SV1
models), SGI PowerChallenge and Origin systems, Sun Enterprise, and
IBM F50 and H70 multiprocessor servers. Parallelization has been
achieved through a careful separation of the code memory into private
(task local) and shared portions. Whenever possible {\sc fortran}
loops have been parallelized over mesh patches which are the entities
requiring the largest amount of work. The most time consuming parts of
the adaption step and patch integration are all parallelized following
the above practice (grey shaded region in Fig.~\ref{f:flow}). \AMRA\
further allows performing calculations with the base level only. In
such a ``non-adaptive'' mode, the adaptive mesh modules of \AMRA\
effectively serve as a domain decomposition tool allowing to partition
the whole computational domain into smaller blocks. From that point of
view, the AMR modules of \AMRA\ offer a very quick and efficient way
for parallelization of the existing hydrocodes. In what follows,
however, we focus solely on the parallel efficiency of the
fully-adaptive code.

\subsubsection{Parallel performance}			\label{s:pperf}

\begin{figure}
\begin{center}
\includegraphics[bb =  94 296 428 629, height=8 cm]{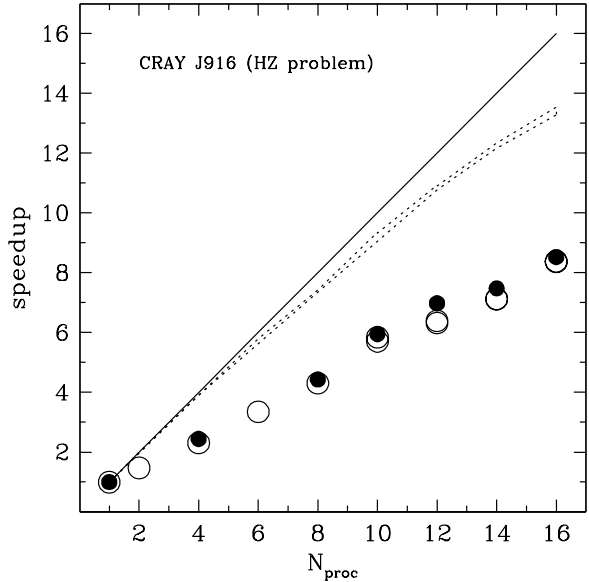}
\caption[]{ 
Parallel performance of \AMRA\ on a CRAY J916 system for the
Hawley-Zabusky problem. Open and filled circles correspond to a small
(patch size $60\times10$) and large (patch size $120\times20$) problem
size. Cray Hardware Performance Monitor results are shown with dotted
lines and indicate average processor usage for a small (lower curve)
and large (upper curve) problem size. The solid line corresponds
to $100\%$ of parallelism.
}
\label{f:speedupHZ}
\end{center}
\end{figure}

Reports of the parallel performance of \AMRA\ obtained with the
ATEXPERT utility on Cray PVP systems showed that for sufficiently
large ($\sim 100$ Mwords) problems the code typically achieves $>
98$\% of parallelism. According to Amdahl's Law \cite{a67} such a
degree of parallelism should offer a theoretical maximum speedup of
about 7 and 12 on 8 and 16 CPUs, respectively. To verify this
prediction we performed several \AMRA\ runs for the Hawley-Zabusky
test (see Sect.~\ref{s:hz}) with a patch size of $60\times10$. We used
a CRAY J916 system running UNICOS 9.0 and Cray CF90 compiler version
2.0.3.1 in non-dedicated mode under very low system load
conditions. The ratio between CPU times for parallel and sequential
runs is shown in Fig.~\ref{f:speedupHZ}
with open circles for a problem of small size (patch size
$60\times10$, effective resolution $960\times160$, average memory use
$\sim 48$ Mwords). The obtained speedup corresponds to $\sim 94$\% of
parallelism. A similar result has been obtained for a 4 times larger
problem (patch size $120\times20$; shown with filled circles in
Fig.~\ref{f:speedupHZ}) indicating that within the measurement errors
the parallel performance of the code does not dependent on the problem
size. The actual speedup (symbols in Fig.~\ref{f:speedupHZ}) is
significantly smaller than the average number of concurrent processors
used during runtime (processor load; dashed lines in
Fig.~\ref{f:speedupHZ}) as reported by the Cray Hardware Performance
Monitor.

\begin{figure}
\begin{center}
\includegraphics[bb =  94 296 428 629, height=8 cm]{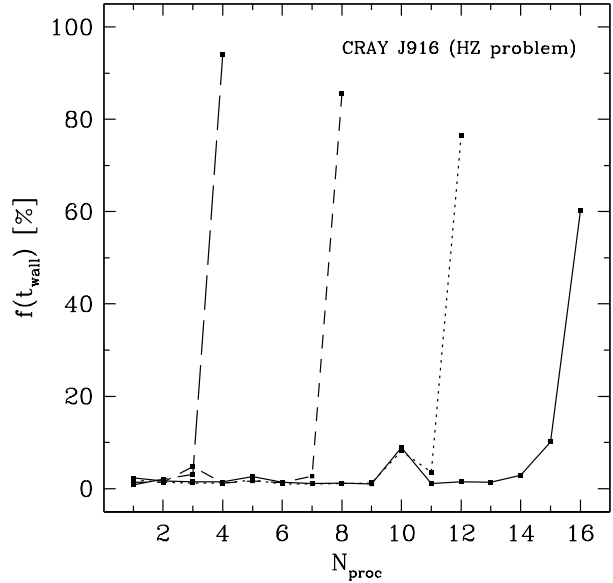}
\caption[]{ 
Fraction of wall-clock time executed with a certain number of
processors for \AMRA\ on a CRAY J916 system for the Hawley-Zabusky
problem. Results for 4, 8, 12, and 16 processor runs are shown (left
to right).
}
\label{f:tconnHZ}
\end{center}
\end{figure}

Fig.~\ref{f:tconnHZ}
shows the fraction of wall-clock time executed using a certain number
of processors. In all runs the code uses the maximum number of
processors most of the time indicating a sufficiently large problem
size. Hence, Fig.~\ref{f:speedupHZ} and \ref{f:tconnHZ} imply that the
large discrepancy between speedup and processor load is due to a large
parallel overhead and a load imbalance rather than due to the presence
of sequential code.

The solution to this problem could be a scheduling scheme for
execution which uses subsets of several patches rather than individual
patches, where each subset requires comparable work. Solvers which use
the directional splitting method need additional communication
between sweeps to exchange boundary information between siblings at
the fine-fine boundaries. This requires frequent synchronization
during single level integration increasing the overall parallel
overhead.

\subsubsection{Performance optimization}		\label{s:perfopt}

Parallel overhead and load imbalance are not the only factors which
determine the performance of \AMRA. As we have already mentioned in
Sect.~\ref{s:data}, the use of extremely small patches might be
disastrous for code performance on machines with a vector
architecture, and may require additional modifications to the solver
(or its complete rewrite) in order to achieve a reasonable
performance. A small patch size significantly increases the AMR
overhead due to the relative increase in the number of operations
required at the patch boundaries (interpolation and conservative fixup
of fluxes). These operations are likely to limit the code performance
on vector machines since they involve fine grained computations and
require indirect remote addressing.

\begin{figure}
\begin{center}
\includegraphics[bb =  84 298 444 629, height=8 cm]{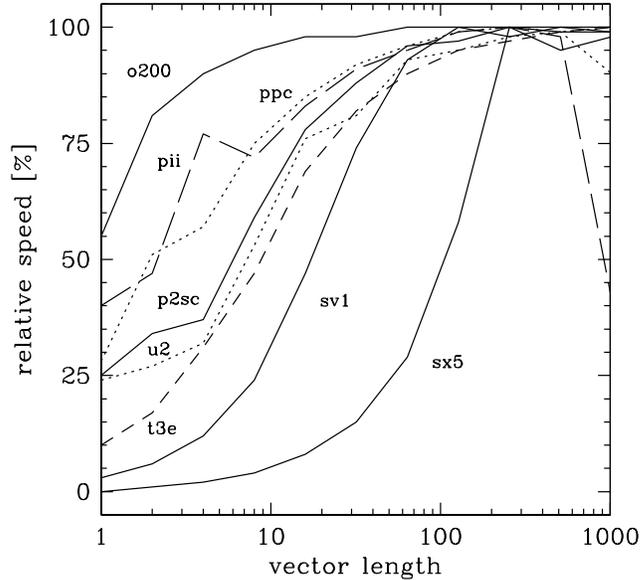}
\caption[]{ 
Relative performance of selected superscalar and vector machines for
the $V1=V2*(A+B)$ operation.  Here $A$ and $B$ are $n\times5$ matrices
and $V1$ and $V2$ are vectors of length $n$. {\bf o200}: SGI
Origin200; {\bf ppc}: IBM PowerPC F50; {\bf pii}: Intel PentiumII;
{\bf p2sc}: IBM P2SC; {\bf u2}: Sun UltraSPARC-II; {\bf t3e}: CRAY T3E
DEC Alpha 21164 (EV5); {\bf sv1}: CRAY SV1-1A; {\bf sx5}: NEC SX-5/3C.
}
\label{f:vperf}
\end{center}
\end{figure}

In order to demonstrate that the optimization of the box sizes is
crucial for achieving reasonable code performance,\footnote{An
optimization step for clustering is routinely done on parallel
machines in order to minimize load imbalance which often determines
the overall performance of parallel codes (see
Sect.~\ref{s:discussion}).} on scalar machines and especially on
machines with vector architecture, we compared the relative
performance for an operation which is typical for the hydrodynamic
advection step. It has the form $V1=V2*(A+B)$. Here $A$ and $B$ are
$n\times5$ matrices (because the number of advected conserved hydrodynamic
quantities in 3-D is 5) matrices and $V1$ and $V2$ are vectors of
length $n$. The results are presented in Fig.~\ref{f:vperf}.
In this test all superscalar architectures achieve (at least) 50\% of
their peak performance for $n=10$ and 75\% for $n=20$. On vector
machines the performance scales almost linearly with the number of
vector elements up to the vector length register (64 for CRAY SV1 and
256 for NEC SX-5). For this reason the performance loss is not as
severe on the SV1 as on the SX-5. A performance level of 50\% is
achieved for vectors not shorter than $\sim 20$ and $\sim 110$ on the
SV1 and the SX-5, respectively. These results are used by the adaption
module (during the optimization step) in form of a cost function for
vector operations which favours merging of small patches.

\section{Results}					\label{s:app}

In the following we present results of the application of \AMRA\ to
several selected one- and two-dimensional flow problems.

\subsection{Two interacting blast waves}		\label{s:bw}

\begin{figure}
\begin{center}
\includegraphics[bb =  53 430 278 686, height=6 cm]{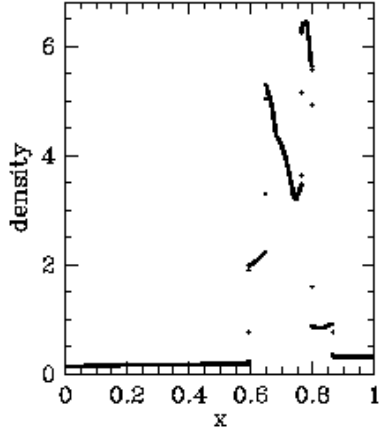}
\caption[]{ 
Solution to the blast waves problem obtained on a single grid with
6400 zones. Shown is the density at time $t=0.038$.
}
\label{f:bwSINGLE}
\end{center}
\end{figure}

The colliding blast waves problem \cite{w82,wc84} is one of the most
demanding tests of hydrodynamic codes and now widely accepted as a
benchmark for newly developed hydrodynamic schemes and their
implementations. The initial conditions for this test are two hot
regions of unequal pressures inside the interval $0{\le}x{\le}1$,
\[
  \vec{U}(0{\le}x{\le}0.1,t=0)
= \pmatrix{ \rho \cr u \cr p    }
= \pmatrix{ 1    \cr 0 \cr 1000 },
\]
and
\[
  \vec{U}(0.9{\le}x{\le}1,t=0)
= \pmatrix{ \rho \cr u \cr p   }
= \pmatrix{ 1    \cr 0 \cr 100 },
\]
separated by a low-pressure cavity,
\[
  \vec{U}(0.1{<}x{<}0.9,t=0)
= \pmatrix{ \rho \cr u \cr p    }
= \pmatrix{ 1    \cr 0 \cr 0.01 }.
\]
The initial data leads to the formation of two shock waves of
unequal strengths which after collision form a weak contact
discontinuity. Fig.~\ref{f:bwSINGLE}
shows the density distribution of a ``converged'' (6400 zones)
single-level run at time $t=0.038$. Two contact discontinuities that
formed at the beginning of the evolution are visible at $x\approx 0.6$
and $x\approx 0.8$, respectively. Another weak contact discontinuity
is visible at $x\approx 0.75$.It formed during collision of the shock
waves propagating to the left ($x\approx 0.65$) and right ($x\approx
0.85$), respectively.

\begin{figure}
\begin{center}
\includegraphics[bb =  51 56 470 686, height=15.2 cm]{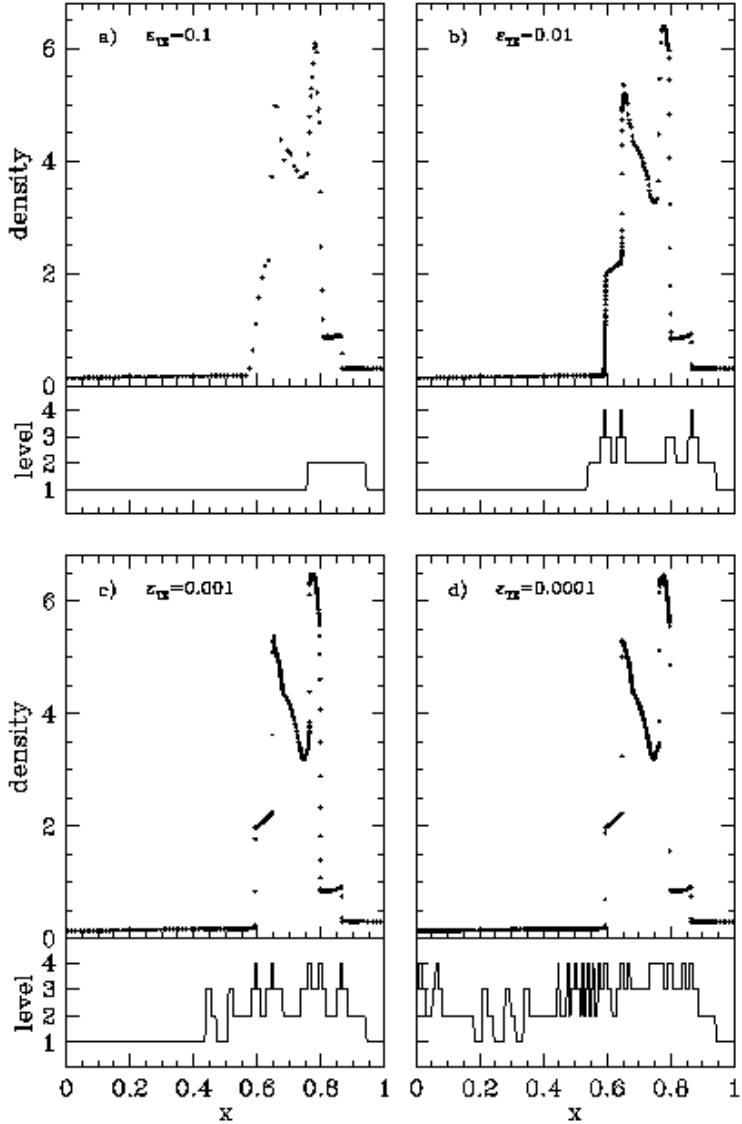}
\caption[]{ 
Solution to the blast waves problem at time $t=0.038$ obtained with
\AMRA. The density in the valid (i.e., not further refined) regions and the
grid level distribution are shown for {\TE}=0.1 (top-left), 0.01
(top-right), 0.001 (bottom-left) and 0.0001 (bottom-right),
respectively.
}
\label{f:bwTE}
\end{center}
\end{figure}

We performed further \AMRA\ runs with an effective resolution equal to
that of the single-level run but using different criteria for
flagging. Firstly, we used the method based on truncation error
estimation with the truncation error threshold, {\TE}, equal to 0.1, 0.01,
0.001, and 0.0001, respectively. From a comparison of the final models
(Fig.~\ref{f:bwTE})
we see that the most important flow features are captured at a
truncation error threshold $\TE=0.001$ (Fig.~\ref{f:bwTE}c) although
the right contact discontinuity was temporarily lost during the
earlier phases of the evolution and was only recaptured later by
additional refinements. Using a still smaller value of {\TE} (0.0001,
Fig.~\ref{f:bwTE}d) helps in resolving all flow features as accurately
as in the single level run. It can be noted that the solution for
$\TE=0.0001$ does not differ much from the result obtained with
$\TE=0.001$. This might be an indication that this solution has been
obtained at a level of accuracy comparable to the truncation error of
the \AMRA\ itself. On the other hand, the use of {\TE} equal to 0.01
or greater degrades the quality of the solution significantly and the
weak contact discontinuity is only barely resolved.

\begin{figure}
\begin{center}
\includegraphics[bb =  55 360 471 687, height=7.75 cm]{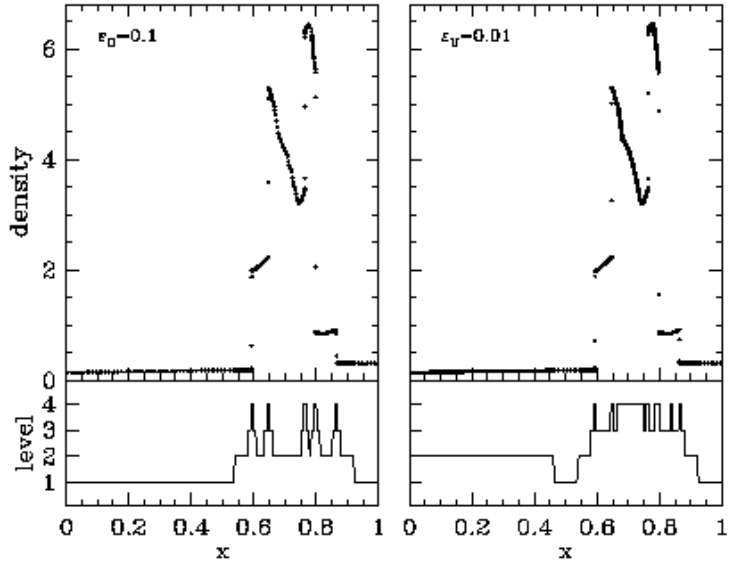}
\caption[]{ 
Solution to the blast waves problem at time $t=0.038$ obtained with
\AMRA. The density in the valid (i.e., not further refined) regions and the
grid level distribution are shown for $\TU{\rho}=\TU{p}=0.1$ and
$\TU{\rho}=\TU{p}=0.01$ in the left and right panel, respectively.
}
\label{f:bwJ}
\end{center}
\end{figure}

Finally, we performed two additional \AMRA\ runs using thresholds for
the relative changes of hydrodynamic variables, \TU{U}, as the only refinement
criteria. The results obtained with density and pressure as indicator
variables with $\TU{\rho}=\TU{p}=0.1$ and $\TU{\rho}=\TU{p}=0.01$
are shown in Fig.~\ref{f:bwJ}.
In both cases the quality of the solution is comparable to the
converged single-level solution. All dominant discontinuities are well
resolved.  The only apparent differences occur in the post-shock
region of the left shock. This flow structure is formed during early
phases of the evolution when the rarefaction propagating initially to
the left is reflected at the left grid boundary and interacts with the
post-shock region of the shock propagating towards the right.

Table~\ref{t:bwPERF}
\begin{table}
\caption[]
{
Performance data for the blast waves problem. The single level run was
performed with 6400 zones.
}
\label{t:bwPERF}
\begin{flushleft}
\begin{tabular}{llrr}
\hline
\noalign{\smallskip}
code      & model                  &    CPU time [s]   &   speedup         \\
\noalign{\smallskip}
\hline
\noalign{\smallskip}
\AMRA     & single level           &        4585       &                   \\
\noalign{\smallskip}
\hline
\noalign{\smallskip}
\AMRA     & $\TE=0.1$              &        70         &    65             \\
\AMRA     & $\TE=0.01$             &       170         &    27             \\
\AMRA     & $\TE=0.001$            &       241         &    19             \\
\AMRA     & $\TE=0.0001$           &       706         &     7             \\
\AMRA     & $\varepsilon_{U}=0.1$  &       157         &    29             \\
\AMRA     & $\varepsilon_{U}=0.01$ &       492         &     9             \\
\noalign{\smallskip}
\hline
\end{tabular}
\end{flushleft}
\end{table}
presents a summary of the performance data (CPU time has been measured
for additional runs with minimal I/O) for the blast waves problem. For
this test problem \AMRA\ offers a speedup between 7 and 19 when
{\TE}-based flagging gave acceptable results. The speedup is even
larger (between 9 and 29) when only the relative changes in density
and pressure are used for flagging.

\subsection{Compact supernova remnant}			\label{s:csnr}

Our first astrophysically relevant problem involves two shock waves
formed due to the interaction between the material ejected by a
supernova explosion and a dense circumstellar medium of constant
density \cite{pr97}. As a result of this interaction a forward and
reverse shock are formed. The shocked material separating the two
shock waves is allowed to cool assuming optically thin conditions. The
radiative losses are calculated explicitly with an equilibrium cooling
function \cite{cloudy,p95}.

\begin{figure}
\begin{center}
\includegraphics[bb =  57 363 578 652, width=\columnwidth]{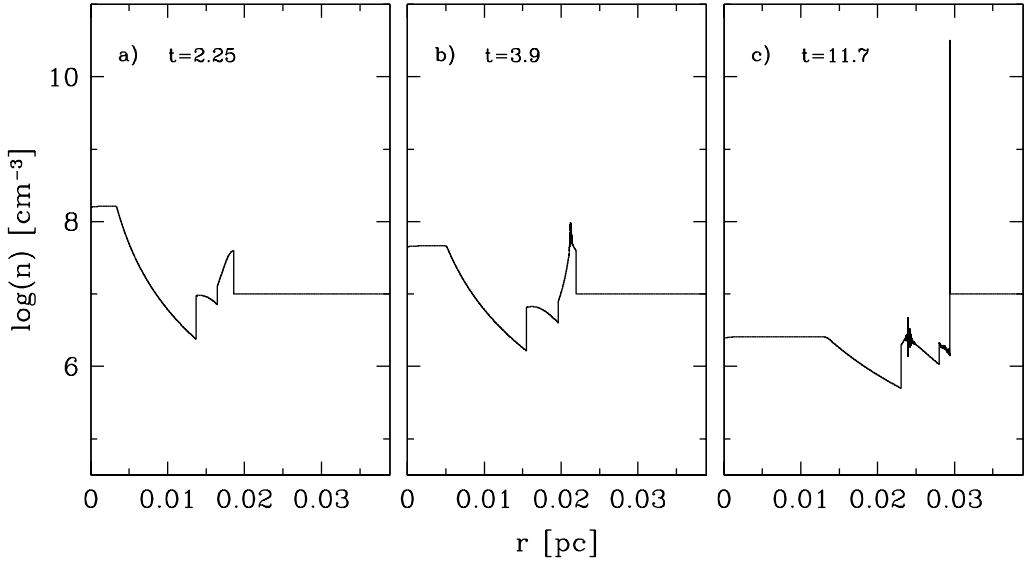}
\caption[]{ 
Solution of the compact supernova problem obtained on a single grid
of 38400 zones. Shown is the logarithm of the number density at
{\bf a)} $t=2.25$\,yr, {\bf b)} $t=3.9$\,yr, and {\bf c)}
$t=11.7$\,yr, respectively.
}
\label{f:csnrSDENS}
\end{center}
\end{figure}

\begin{figure}
\begin{center}
\includegraphics[bb =  132 352 441 640, height=8 cm]{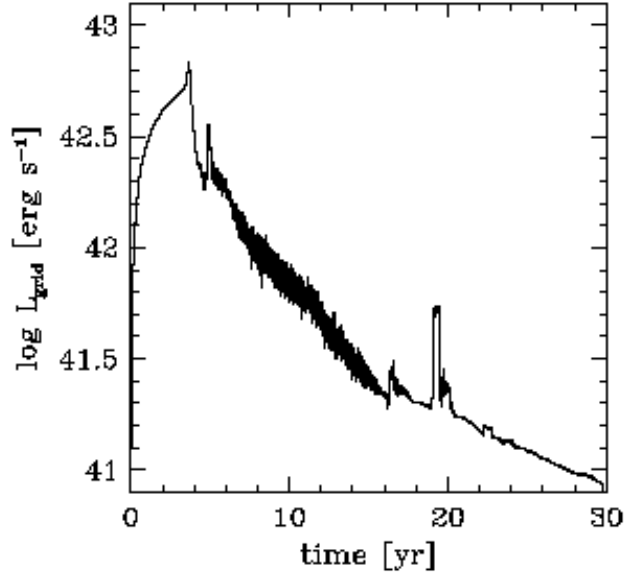}
\caption[]{ 
Temporal evolution of the luminosity for a supernova remnant evolving
in a dense ($n=10^7$\,cm$^{-3}$) medium computed on a single grid with
a resolution of 38400 zones.
}
\label{f:csnrSLUM}
\end{center}
\end{figure}

The density profile obtained with a single-level grid at
a resolution of 38400 zones is shown in Fig.~\ref{f:csnrSDENS}.
Early in the evolution the supernova ejecta drive a forward shock into
the ambient medium while the reverse shock propagates into the
unshocked ejecta (Fig.~\ref{f:csnrSDENS}a). After approximately one
cooling time the loss of pressure due to cooling in the
post-forward-shock region becomes significant leading to the formation
of pressure gradients which slowly accelerate the gas towards the low
pressure region. Once the gas temperature in this region drops below
$\simeq 2\times10^7$\,K the emissivity of the gas begins to rise with
decreasing temperature eventually resulting in a ``catastrophic
cooling'' \cite{f81,t+92} followed by a relatively short phase of
rapid mass accumulation in a dense shell. Fig.~\ref{f:csnrSDENS}b
shows the density structure soon after catastrophic cooling occurred.
A similar sequence of events also takes place behind the reverse shock
(Fig.~\ref{f:csnrSDENS}c) where catastrophic cooling occurs at a later
time ($t\approx12$ yr) due to the lower densities in this region. The
formation of the forward shell is accompanied by a rapid increase of
the total luminosity around time $t=3.8$ yr (Fig.~\ref{f:csnrSLUM}),
while the formation of the reverse shell manifests itself as a change
of slope of the light curve around $t=11.5$ yr.

Since the minimum temperature of the gas allowed in our simulation
($10^4$ K) is typically 3 to 4 orders of magnitude lower than the
post-shock temperature, the cold shell appears as a very thin
structure which is extremely difficult to resolve on a single grid. We
found that in order to resolve the shell a resolution of at least
$\sim 10^5$ equidistant zones has to be used \cite{p95}. Such a
simulation would be prohibitively expensive. Also, since the most
important structures (the two shocks and their dense shells) occupy
only a small fraction of the total volume, very high speedups (several
hundred) are to be expected applying adaptive grid techniques.

In our \AMRA\ test run we kept the refinement criteria fixed while the
number of levels, i.e. the effective resolution, has been increased
between the runs. We used truncation error estimation with $\TE=0.01$,
flagging relative changes in density ($\TU{\rho}=1$) and total energy
($\TU{{\rho}E}=1$). Shocks were flagged with $\TU{p}=1$ and contact
discontinuities with $\TU{\rho}=0.1$. In addition, we decided to
unflag (coarsen) the inner part of the computational grid from the
center out to the (time-dependent) radius of the reverse shock.

\begin{figure}
\begin{center}
\includegraphics[bb =  57 293 578 652, width=\columnwidth]{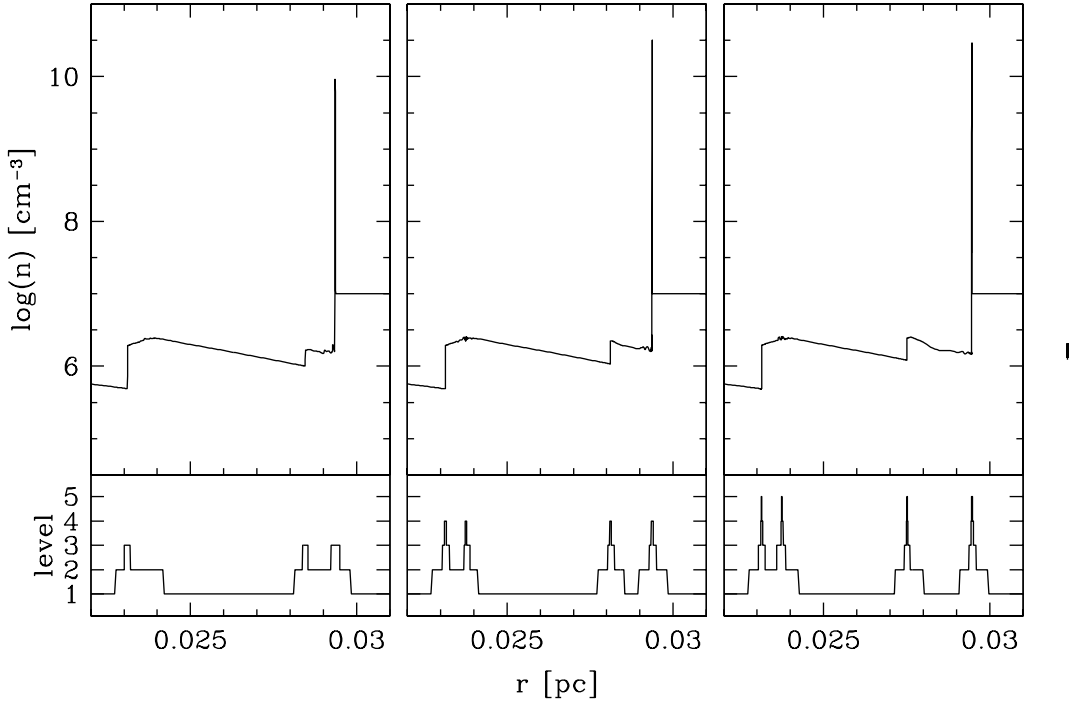}
\caption[]{ 
Solution to the compact supernova problem at $t=11.7$ yr obtained with
\AMRA. Shown are the logarithm of the number density in the valid
(i.e., not further refined) regions and the grid level distribution:
{\bf a)} 3 levels, {\bf b)} 4 levels, {\bf c)} 5 levels.
}
\label{f:csnrDENS}
\end{center}
\end{figure}

We used three, four and five grid levels with an effective resolution of
9600, 38400, and 153600 zones, respectively. Therefore, the medium
resolution \AMRA\ run had an effective resolution equal to that of the
single-level run. Fig.~\ref{f:csnrDENS}
shows the density profiles at $t=11.7$ yr obtained with \AMRA. In all
three cases the result closely resembles that obtained with a single
grid (Fig.~\ref{f:csnrSDENS}). The density decreases from the center
through the unshocked ejecta up to the reverse shock at $r\approx
0.023$ pc. The density spike located right behind the reverse shock
marks the position at which a secondary shell is just
forming. Adjacent to the density spike is region of hot gas with
linearly decreasing density which is further reheated by a weak
secondary shock near its rightmost end. The density in the primary
shell ($r\approx 0.029$ pc) is equal to $\approx 10^{11}$\,cm$^{-3}$,
except for the lowest resolution \AMRA\ model. The post-shock region
of the forward supernova shock located just outside the primary shell
remains unresolved in all but the highest resolution \AMRA\ run. Only
then which the resolution is sufficiently high to observe the
oscillatory instability of the cooling shock \cite{ci82,p93,wf96}.

Comparing CPU times of the single-grid and AMR run obtained at the
same effective resolution (Table~\ref{t:csnrPERF})
\begin{table}
\caption[]
{
Performance data for the compact supernova problem.
}
\label{t:csnrPERF}
\begin{flushleft}
\begin{tabular}{lrrc}
\hline
\noalign{\smallskip}
model          & resolution &    CPU time [s]   &       speedup         \\
\noalign{\smallskip}
\hline
\noalign{\smallskip}
\AMRA\ 1 level  &     9600   &       23260       &                       \\
\AMRA\ 1 level  &    38400   &      430300       &                       \\
\noalign{\smallskip}
\hline
\noalign{\smallskip}
\AMRA\ 3 levels &     9600   &         851       &        27             \\
\AMRA\ 4 levels &    38400   &        3571       &       120             \\
\AMRA\ 5 levels &   153600   &       22690       &       233$^{\rm a}$   \\
\noalign{\smallskip}
\hline
\end{tabular}
\begin{list}{}{}
\item[$^{\rm a}$] Estimated.
\end{list}
\end{flushleft}
\end{table}
indicates a speedup of about 27 and 120 for 3 and 4 level AMR runs,
respectively. From the total number of zones to be updated during the
single grid run and from the execution speed measured in number of
zones updated per second, we estimate that at a resolution of 153600
zones the speedup would exceed 200.

\subsection{Hawley-Zabusky problem}			\label{s:hz}

Hawley and Zabusky \cite{hz89} studied numerically the interaction
between an oblique shock and a contact discontinuity. At the beginning
of the evolution a shock tube is filled with a gas at rest containing
a contact discontinuity inclined at a small angle (30$^\circ$) with
respect to the front of a Mach 1.2 planar shock wave. Vorticity
deposition which occurs during the passage of the shock wave through
the contact discontinuity leads to the formation of vortices which
interact and subsequently merge.

The amount of vorticity deposited at the contact discontinuity
sensitively depends on the numerical resolution and internal
dissipation of the advection scheme. At late times the rollup of
vortices is additionally affected by the ``far field'' produced by
sound waves generated during the early interaction of the shock wave
with the contact discontinuity and to some degree also by weak waves
reflected from both ends of the shock tube. For these reasons no
numerically ``converged'' solution can be obtained. The final
structure (location and number of vortices) changes with grid
resolution, and is extremely sensitive to inaccuracies introduced by
any numerical scheme.

\begin{figure}
\begin{center}
\includegraphics[bb =  221 358 374 483, width=\columnwidth]{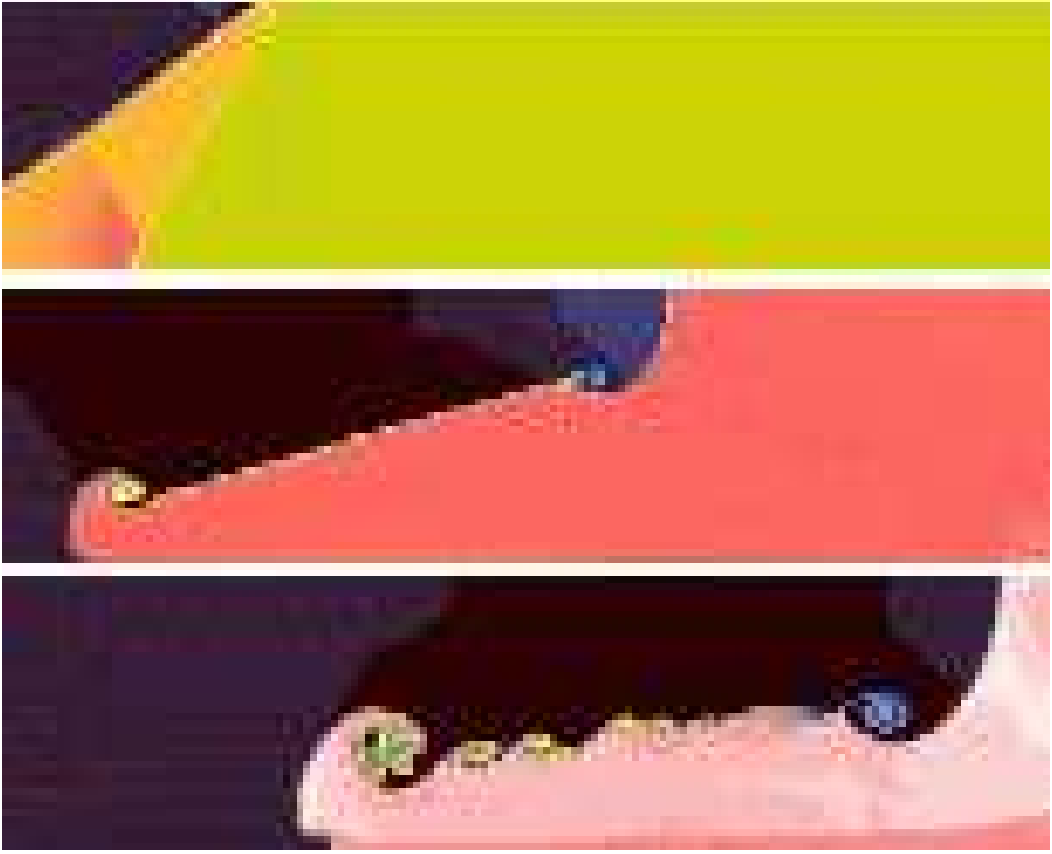}
\caption[]{ 
Solution to the Hawley-Zabusky problem obtained on a single grid
(resolution $960\times160$ zones). The density is shown at times $t=91$
(top), $t=364$ (middle), and $t=620$ (bottom), respectively.
}
\label{f:hzSDENS}
\end{center}
\end{figure}

\begin{figure}
\begin{center}
\includegraphics[bb =  66 203 482 629, width=0.8\columnwidth]{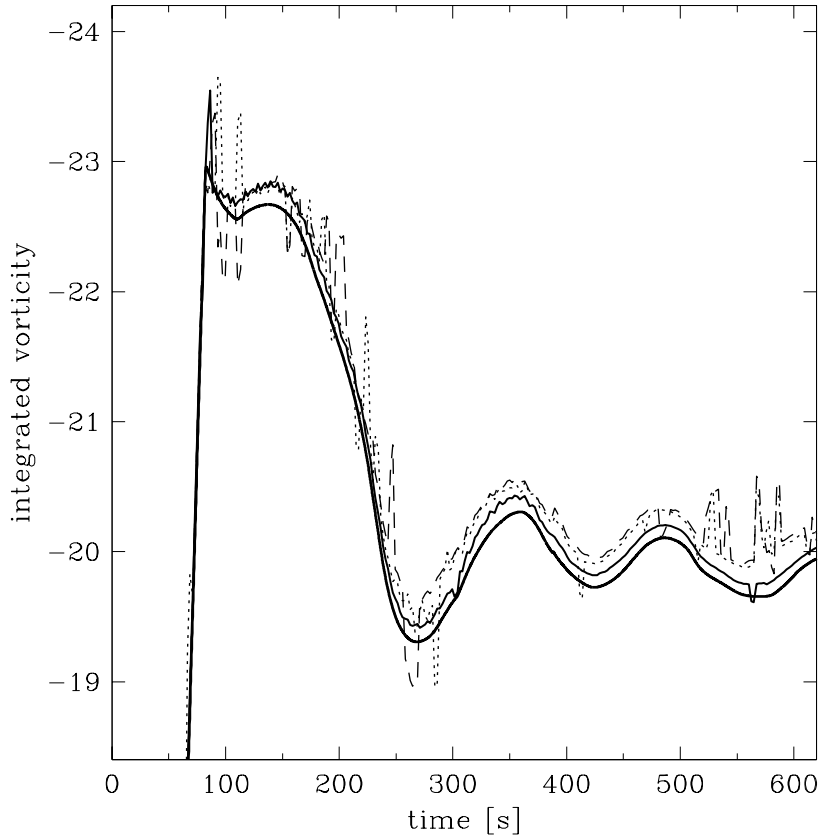}
\caption[]{ 
Temporal evolution of the integrated vorticity in the Hawley-Zabusky
test. Single grid model: thick solid line. \AMRA\ models: $\TE=0.1$
(dashed), $\TE=0.01$ (dotted), $\TE=0.001$ (thin solid).
}
\label{f:hzVORT}
\end{center}
\end{figure}

Fig.~\ref{f:hzSDENS}
shows a sequence of images obtained on a single grid with a resolution
of $960\times160$ zones. The integrated vorticity reaches an absolute
maximum (solid line in Fig.~\ref{f:hzVORT})
at $t=91$ when the shock wave passes the full horizontal extent of the
contact discontinuity (Fig.~\ref{f:hzSDENS}a). During the following evolution
vortices first grow at the smallest resolved scales ($t=364$,
Fig.~\ref{f:hzSDENS}b). Later they interact and merge with only a few
large vortices remaining at the final time ($t=620$,
Fig.~\ref{f:hzSDENS}c).

\begin{figure}
\begin{center}
\includegraphics[bb =  221 358 374 483, width=\columnwidth]{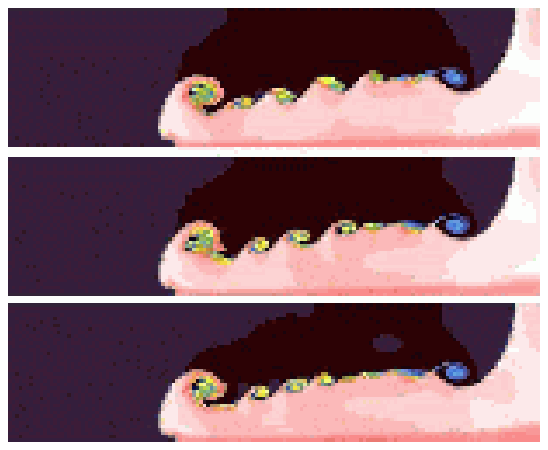}
\caption[]{ 
Solution to the Hawley-Zabusky problem obtained with \AMRA. The
density in the valid (i.e., not further refined) regions is shown at $t=620$
and $\TE=0.1$ (top), $\TE=0.01$ (middle), and $\TE=0.001$ (bottom),
respectively.
}
\label{f:hzDENS}
\end{center}
\end{figure}

\AMRA\ simulations have been performed at the same effective
resolution but changing the truncation error threshold for zone
flagging. To keep the initial discontinuities fully resolved we also
used $\TU{\rho}=1$ and $\TU{p}=0.5$. The former criterion helps in
resolving the contact discontinuity while the latter ensures creation
of the finest level patches near the shock front. \AMRA\ models at the
final time obtained with \TE\ equal to 0.1, 0.01, and 0.001 are shown
in the upper, middle, and lower part in Fig.~\ref{f:hzDENS},
respectively. It can be seen that vortices are progressively more
diffused with increasing \TE. This indicates that the \TE\ criterion
can be used to control the amount of numerical diffusion present in
\AMRA\ simulations.

The temporal evolution of the integrated vorticity is shown in
Fig.~\ref{f:hzVORT} with thin solid, dotted, and dashed lines for \TE\
equal to 0.1, 0.01, and 0.001, respectively. In all cases low
amplitude wiggles (clearly visible at maximum and around $t\approx
350$ in model $\TE=0.001$) or sudden erratic changes (visible for
$\TE=0.001$ around $t\approx 310$ and $t\approx 570$) in the
integrated vorticity can be observed. In addition, large amplitude
variations of the total vorticity are observed with a net variation
close to zero. We identify the former phenomenon with the destruction
and subsequent recreation of patches in localized regions of the
computational domain which contain substantial amount of vorticity. A
remedy for this problem might be a delayed patch destruction
\cite{nirvana,flash} or the direct introduction of a temporal
smoothing into the flagging procedure.

Finally, following Quirk \cite{q91}, we note that any AMR patch corner
acts as a potential obstacle for the flow and becomes a source of
spurious vorticity once the flow is not exactly aligned with one of
the coordinate directions. This observation may explain why the
vorticity seems to differ slightly from that of the single-grid run
already prior to reaching its maximum (although until this moment both
the shock front and the contact discontinuity are always covered with
finest level patches). The amount of spurious vorticity is likely to
be higher in simulations performed with larger refinement
ratios. These issues should be taken into account when performing AMR
simulations for problems involving convection or turbulence (see
\cite{c99} for a discussion of problems related to modelling turbulent
flows in the framework of large eddy simulations).

\begin{figure}
\begin{center}
\includegraphics[bb =  145 299 451 543, width=\columnwidth]{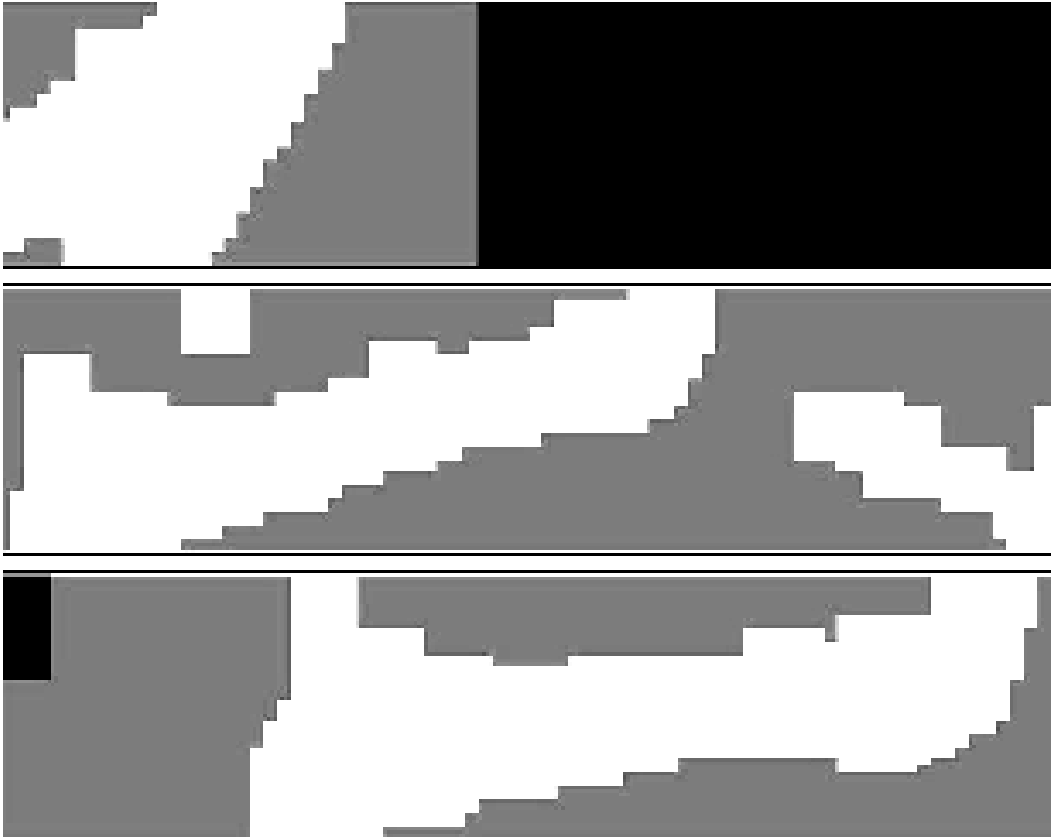}
\caption[]{ 
Distribution of grid levels in \AMRA\ run for the Hawley-Zabusky
problem with $\TE=0.001$. Regions covered with the base level are
shown in black and the finest level patches are shown in white: {\bf a)}
$t=93$, {\bf b)} $t=365$, {\bf c)} $t=620$.
}
\label{f:hz3LEV}
\end{center}
\end{figure}

Table~\ref{t:hzPERF}
\begin{table}
\caption[]
{
Performance data for the Hawley-Zabusky problem.
}
\label{t:hzPERF}
\begin{flushleft}
\begin{tabular}{llll}
\hline
\noalign{\smallskip}
code      & model              &    CPU time [s]   &       speedup     \\
\noalign{\smallskip}
\hline
\noalign{\smallskip}
\AMRA\ 1 level & $960\times160$     &       53070       &              \\
\noalign{\smallskip}
\hline
\noalign{\smallskip}
\AMRA         & $\TE=0.1$          &       14540       &        3.6    \\
\AMRA         & $\TE=0.01$         &       16890       &        3.1    \\
\AMRA         & $\TE=0.001$        &       24890       &        2.1    \\
\noalign{\smallskip}
\hline
\end{tabular}
\end{flushleft}
\end{table}
summarizes the performance data for the Hawley-Zabusky problem.  The
obtained speedups are rather disappointing ranging from 2 to less than
4. This result follows from the fact that a relatively large fraction
of the computational domain is occupied by discontinuities and has to
be resolved at the finest level (Fig.~\ref{f:hz3LEV}). This fraction
starts growing soon after the shock begins to interact with the
contact discontinuity.  At $t\approx90$ (Fig.~\ref{f:hz3LEV}a) the second and
third level occupy about 45\% and 23\% of the domain, respectively.
This trend continues through the middle of the evolution and at
$t=365$ (Fig.~\ref{f:hz3LEV}b) the filling factor for the second and
third level $\sim 96$\% and $\sim 39$\%, respectively. Since the shock
is the only discontinuity escaping the grid the situation does not
change much for a better near the end of the evolution
(Fig.~\ref{f:hz3LEV}c) when the second and third level still cover
$\sim 82$\% and $\sim 29$\% of the domain, respectively. We may
define the maximum speedup as the ratio of the number of zones to be
updated on the finest level (assuming no overlap between sibling
patches) to that in the single grid run. For the Hawley-Zabusky test
and with $\TE=0.001$ this number is equal to $\sim 3.3$ which has to
be compared to the measured speedup of 2.1. The difference between the
two numbers accounts for the CPU time spent integrating all coarser
levels, the overhead due to additional data motion, grid adaption
including error estimation, and operations at the patch boundaries.
These steps account for 3\%, 2\%, 6\% and 17\% of the CPU time in
the \AMRA\ run on a SGI Origin200 system.

\subsection{Supernova shock instability}		\label{s:ers}

The problem of the supernova shock propagation through the atmosphere of
the stellar progenitor has been the subject of several detailed
numerical studies \cite[and references therein]{k+00} following
observational evidence for mixing that must have occurred during the
explosion of supernova SN~1987\,A (for a review see \cite{m98}).

In our \AMRA\ study we used the initial model of M\"uller, Fryxell and
Arnett \cite{mfa91}. Calculations have been performed assuming
spherical symmetry on a grid extending from $r_{\mathrm{in}}=0$ cm to
$r_{\mathrm{out}}=3\times10^{12}$ cm in radius and from $\theta=0$ to
$\theta=\pi/2$ in polar angle. The base level consisted of 4 patches
of $48\times25$ zones allowing for free outflow along the outer radial
grid boundary; at all other boundaries we imposed reflecting boundary
conditions. We used three levels of refinement with a twofold increase
in resolution on the second level in each direction, and refinement
ratios of 4 and 2 in radius and angle for the remaining two
levels. This set-up results in an effective resolution of
$3072\times400$ zones. We used $\TE=0.001$ and $\TU{\rho}=1$ as
refinement criteria. A simple ideal gas equation of state with
$\gamma=4/3$ has been used.  The simulation has been started 300
seconds after the explosion and was followed up to $t=13\,000$
seconds.

\begin{figure}
\begin{center}
\includegraphics[bb = 0 0 566 566, width=\columnwidth]{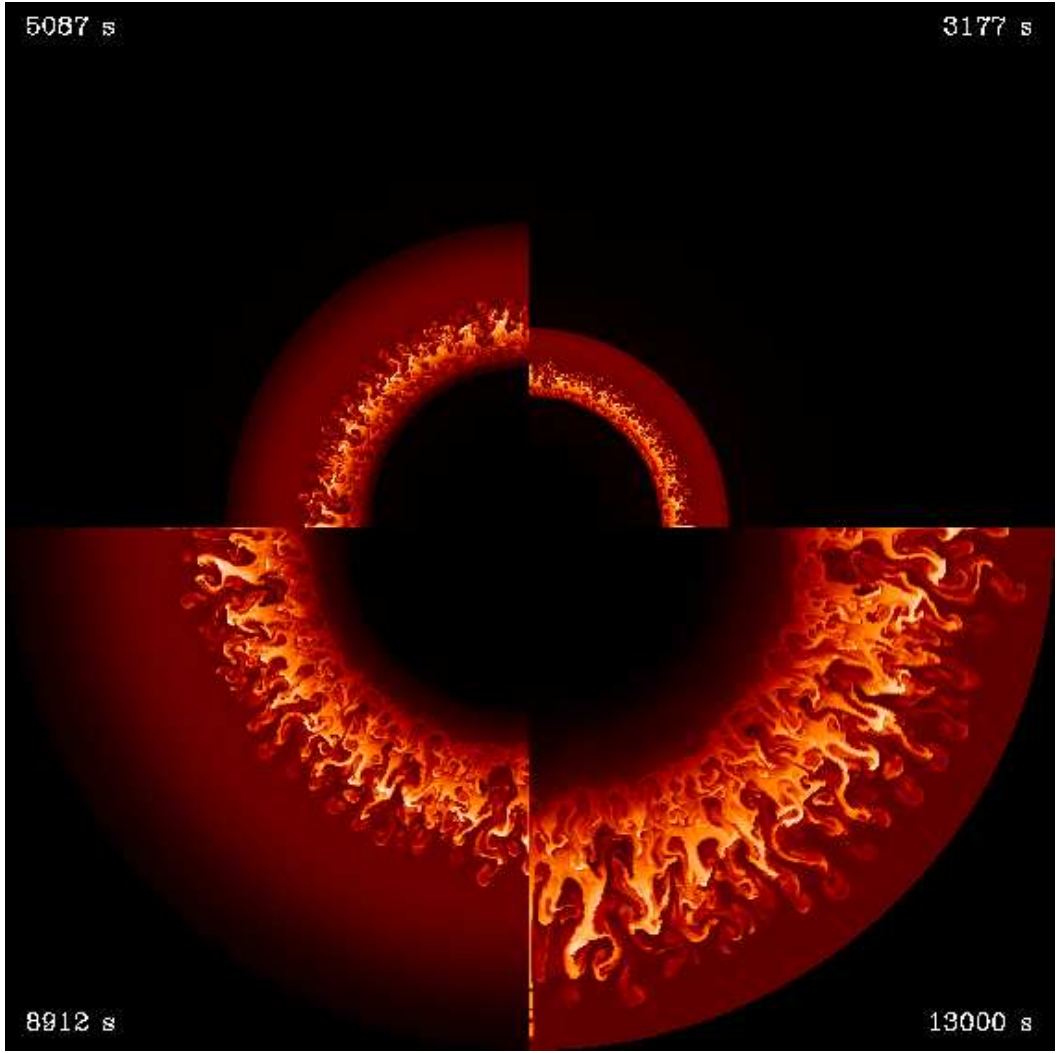}
\caption[]{
Density distribution in \AMRA\ model for the supernova shock
instability problem. The density in each frame has been normalized to
its maximum value:
$1.51\times10^{-2}$ g\,cm$^{-3}$ ($t=3177$),
$5.10\times10^{-3}$ g\,cm$^{-3}$ ($t=5087$),
$1.14\times10^{-3}$ g\,cm$^{-3}$ ($t=8912$),
$4.13\times10^{-4}$ g\,cm$^{-3}$ ($t=13000$).
}
\label{f:ersDENS}
\end{center}
\end{figure}

\begin{figure}
\begin{center}
\includegraphics[bb = 511 721 75 14, width=8cm, angle=90]{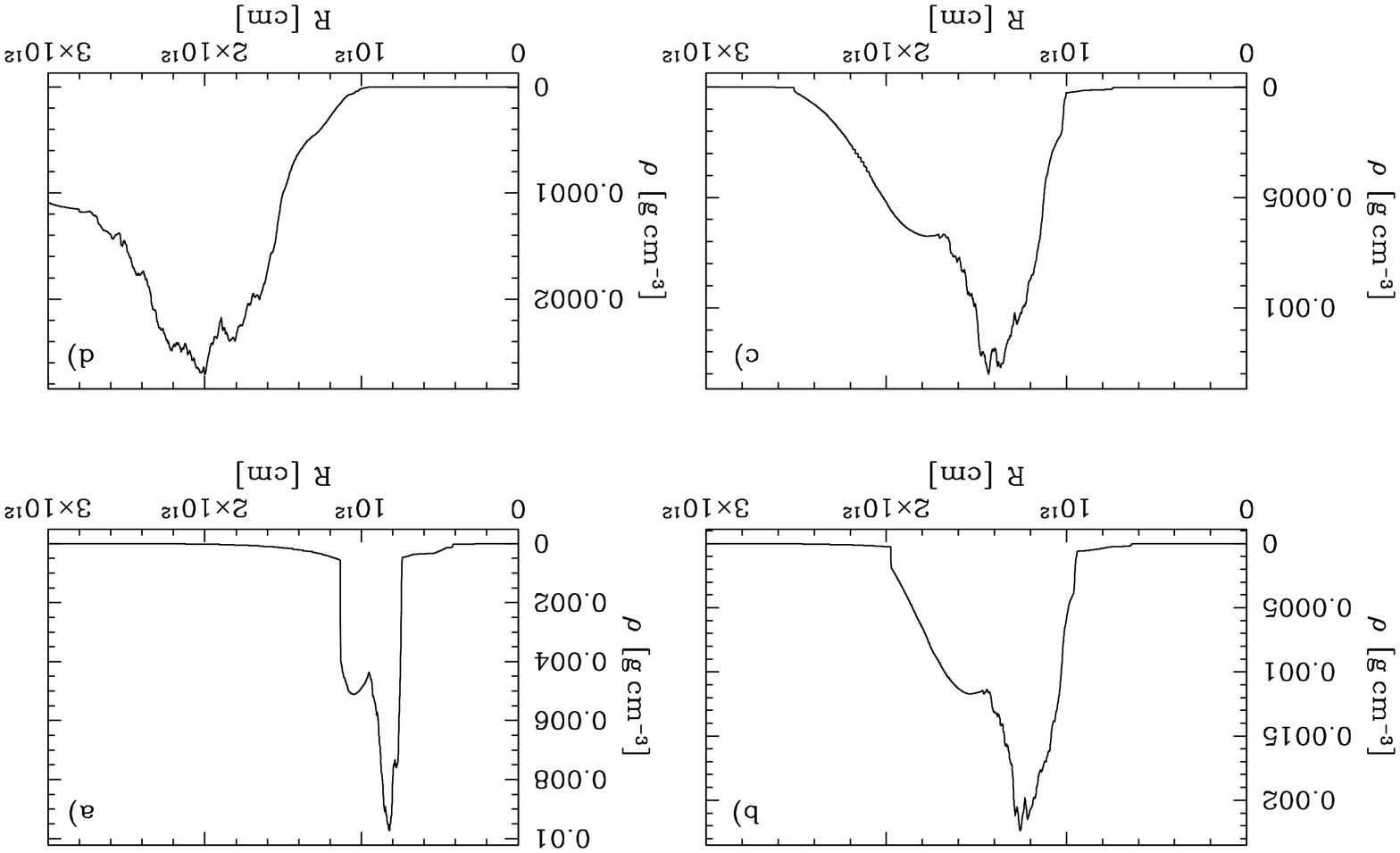}
\caption[]{
Density profile averaged over angle in \AMRA\ model for the supernova
shock instability problem. {\bf a)} $t=3177$, {\bf b)} $t=5745$,
{\bf c)} $t=7008$, {\bf d)} $t=12701$.
}
\label{f:ersMIXL}
\end{center}
\end{figure}

Fig.~\ref{f:ersDENS}
presents the distribution of the gas density between $t=3177$\,s and
the final time. During that period instabilities associated with the
hydrogen-helium (outermost family of mushrooms) and helium-carbon
composition interfaces (more dense mushrooms visible near the middle
of the strongly mixed layer) are already well developed. This result
is in qualitative agreement with the result obtained by M\"uller
\etal\ \cite[Fig.~4]{mfa91} although \AMRA\ predicts the instability
to develop from a shorter (by a factor of $\sim 2$)
wavelength. Moreover, there is also a third instability visible in
\AMRA\ developing in a region traversed by the reverse shock at the
base of the expanding envelope. Most of these differences can be
attributed to the higher (by a factor of $\sim5$) angular resolution
offered by \AMRA\ during the initial stages of the instability
growth. In both models, however, the instability is strong and
produces a ``mixed'' layer of similar average thickness and density
(compare Fig.~\ref{f:ersMIXL}
and Fig.~6 of M\"uller \etal).

\begin{figure}
\begin{center}
\includegraphics[bb = 0 0 566 566, width=\columnwidth]{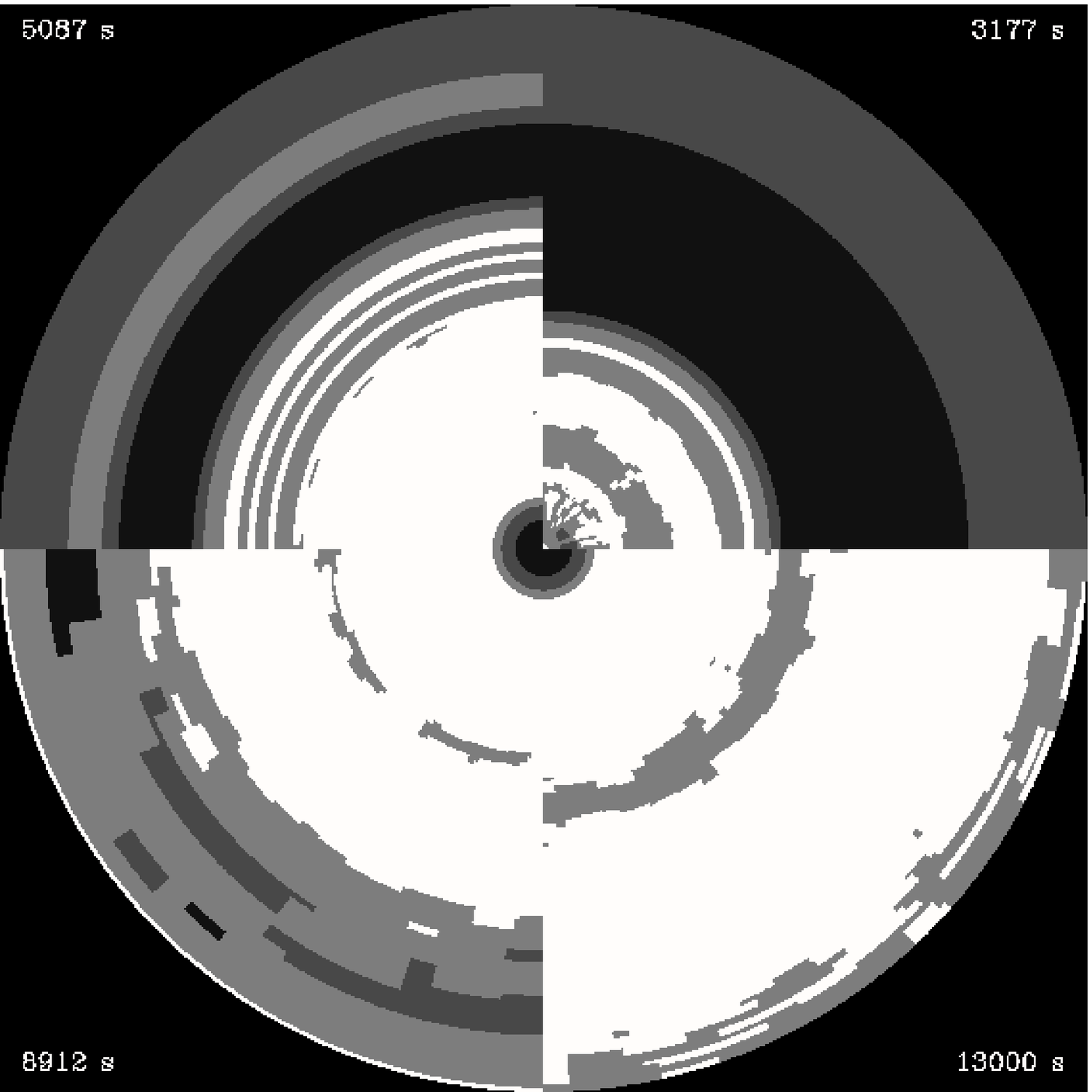}
\caption[]{
Distribution of grid levels in \AMRA\ model for the supernova shock
instability problem. Times in each frame correspond to that of
Fig.~\ref{f:ersDENS}. Base level patches are shown in black and the
finest level regions are shown in white.
}
\label{f:ersLEV}
\end{center}
\end{figure}

Since the finest level occupies a constantly growing
fraction of the total volume (Fig.~\ref{f:ersLEV}),
the speedup obtained in \AMRA\ is largest during early times ($\sim6$
for $t < 3500$) with an average value for the whole run of $\sim
4$. However, we note that in our \AMRA\ run it was possible to allow
the code to run at much ($\sim 10$) larger time steps by decreasing
the resolution in the innermost 10\% of the grid, that is, in the
region which imposes the stringent limits on the time step. In the
\AMRA\ run this region is covered only by the base level for $t >
3500$. To allow for similarly large time steps in a run with uniform
grid one would need to removed the inner part of the star.

\section{Discussion}					\label{s:discussion}

The implementation details of a newly developed Adaptive Mesh
Refinement code, \AMRA, have been presented together with one- and
two-dimensional tests cases. Our experience shows that for a given
resolution the AMR technique offers savings in terms of computer time
as long as the average (over the whole computation) fraction of the
computational domain to be resolved at the finest level (filling
factor) does not exceed about one half of the total volume. This
statement is valid for simple purely hydrodynamic problems. More
substantial savings might be expected for problems involving
computationally expensive source terms.

Although the fractional volume occupied by the finest level is the
major factor determining the overall efficiency of \AMRA, speedups
obtained for the same problem show a strong dependence on processor
architecture (RISC/cache, vector). Our experience on shared memory
parallel systems indicates that the parallel overhead (communication,
synchronization) is significant, and its reduction is likely to be
even more important on systems with distributed memory.

Care has to be taken when simulating flows for which rotation and
shear are important since patch corners effectively act as a source of
spurious vorticity especially when the change of resolution between
levels is substantial. Finally, we note that the AMR approach is not
suitable for problems which require uniform high-resolution. These
include simulations of convective or turbulent flows.

\ack

We thank Micha{\l} R\'o\.zyczka for his ongoing support and interest
in this project. Konstantinos Kifonidis and Pawe{\l} Cieciel\c{a}g
made an enormous contribution to this work by sharing with us their
impressions following ``real world'' applications of \AMRA\ -- we
thank them warmly for being patient with us! The work of TP was partly
supported by the grants 2-1213-91-01, 2.P304.017.07, 2.P03D.004.13,
and 2.P03D.014.19 from the Polish Committee for Scientific Research,
and ESO grant A-01-063. The simulations have been performed on the
CRAY SV1-1A at the Interdisciplinary Centre for Computational
Modelling in Warsaw, the SGI Origin200 at the Nicolaus Copernicus
Astronomical Center, Warsaw, and the CRAY J916/512 at the
Rechenzentrum Garching.


%
%
%
\end{document}